\documentclass[final,3p,times]{elsarticle}

\usepackage{amssymb}
\usepackage{amsmath}
\usepackage{xcolor}
\usepackage{bm}
\usepackage{amsthm}
\usepackage{float}
\usepackage{makecell}
\usepackage{booktabs,multirow}
\usepackage{algorithm}
\usepackage{algpseudocode}
\usepackage{comment}
\usepackage{subfig}
\usepackage{hyperref}
\hypersetup{
  colorlinks   = true, %Colours links instead of ugly boxes
  urlcolor     = blue, %Colour for external hyperlinks
  linkcolor    = blue, %Colour of internal links
  citecolor   = red %Colour of citations
}

\allowdisplaybreaks

\journal{High Impact}

\begin{document}

\begin{frontmatter}



\title{Control Co-Design Under Uncertainty for Offshore Wind Farms: Optimizing Grid Integration, Energy Storage, and Market Participation}
% {Offshore Wind Farm Control Co-Design with Energy Storage under Uncertainty for Market Participation and Grid Frequency Support}
% Control co-desing of Offshore wind farm with energy storage 
% under uncertainty for energy market participation and grid frequency support.}


\author[]{Himanshu Sharma\corref{cor1}} %% Author name
\author[]{Wei Wang} %% Author name
\author[]{Bowen Huang} %% Author name
\author[]{Buxin She} %% Author name
\author[]{Thiagarajan Ramachandaran} %% Author name

\affiliation{organization={Pacific Northwest National Laboratory},%Department and Organization
            city={Richland},
            state={WA},
            country={USA}}

%% Abstract
\begin{abstract}
Offshore wind farms (OWFs) are set to significantly contribute to global decarbonization efforts. Developers often use a sequential approach to optimize design variables and market participation for grid-integrated offshore wind farms. However, this method can lead to sub-optimal system performance, and uncertainties associated with renewable resources are often overlooked in decision-making. This paper proposes a control co-design approach, optimizing design and control decisions for integrating OWFs into the power grid while considering energy market and primary frequency market participation. Additionally, we introduce optimal sizing solutions for energy storage systems deployed onshore to enhance revenue for OWF developers over time. This framework addresses uncertainties related to wind resources and energy prices. We analyze five U.S. west-coast offshore wind farm locations and potential interconnection points, as identified by the Bureau of Ocean Energy Management (BOEM). Results show that optimized control co-design solutions can increase market revenue by 3.2\% and provide flexibility in managing wind resource uncertainties.
\end{abstract}


% \begin{highlights}
% \item  Optimization for the design and control of offshore wind farms for energy market participation.
% \item Multi-stage, multi-timescale stochastic control co-design optimization formulation.
% \item Offshore wind farm and energy storage droop control tuning for primary frequency market participation.
% \item Validation scheme for droop control parameters using EMT-based transient simulations to ensure compliance with reserve requirements.
% \end{highlights}

% %% Keywords
% \begin{keyword}
% control co-design \sep 
% %% keywords here, in the form: keyword \sep keyword

% %% PACS codes here, in the form: \PACS code \sep code

% %% MSC codes here, in the form: \MSC code \sep code
% %% or \MSC[2008] code \sep code (2000 is the default)

% \end{keyword}

\end{frontmatter}


\section{Introduction}\label{sec1}
Offshore wind farms (OWFs) are becoming a key element in the global shift towards sustainable energy. These installations harness the wind over open seas, benefiting from higher wind speeds, which has driven their rapid international expansion. Over the past two decades, the offshore wind industry has achieved significant technological advancements, such as floating platform turbines and efficient energy generation systems. These innovations have increased wind farm capacity, efficiency, and expanded their reach to deeper waters. The 2024 Offshore Wind Market Report \cite{OffshoreWindReport24} indicates the U.S. offshore wind energy pipeline grew by 53\% from the previous year, with notable expansion along the East and West Coasts.

As OWF development and grid integration progress, enhancing grid capacity in the U.S. is crucial \cite{sharma2023grid,brinkman2024atlantic}. Robust transmission infrastructure is essential to transport electricity from offshore sites. Innovations in transmission technologies, like High-Voltage Direct Current (HVDC) cables, promise efficient long-distance electricity transfer. HVDC systems have lower energy losses compared to Alternating Current (AC) systems, making them suitable for wind farms situated far from the coast. Voltage Source Converters (VSC) \cite{Flourentzou2009VSCBasedHP} and Modular Multilevel Converters (MMC) \cite{Zhang2017ModelingCA} are also important for designing multi-terminal HVDC (MTDC) systems for wind farms.

Designing and operating the MTDC network is critical for efficient OWF integration. The MTDC network comprises offshore substations, electrical cables, and potentially onshore power conversion infrastructure. These components can significantly impact overall system costs. Our study focuses on cable size design to maintain problem tractability. Assessing cable size, taking into account operational strategy—such as the ability to meet power requirements at the point of common coupling (PCC)—can prevent under-sizing and over-sizing of transmission cables, which directly influences MTDC network costs.

Beyond MTDC network design and cable sizing, ensuring a stable system is vital. Integrating correctly sized energy storage systems (ESS) with operational controllers provides operational flexibility and improved dispatchability \cite{zhao2023grid,sharma2023grid,luo2015coordinated}. The role of ESS in wind energy resources is reviewed in \cite{Zhao2015ReviewOE}, examining design and control for wind energy. Offshore applications benefit from onshore ESS deployment due to economic and grid service reasons, as detailed in \cite{arellano2022energy,rabanal2024energy}. In our study, we consider ESS at the onshore point of interconnection.

As renewables grow and conventional generators decline, grid control requirements are evolving. OWFs must meet strict grid codes and provide ancillary services to ensure grid stability \cite{ela2012effective,federal2018essential}. Market volatility impacts renewables' participation in electricity markets \cite{eldridge2022modeling,seel2018impacts}. As the market matures, larger projects enjoy lower costs and enhanced profitability. OWFs can explore new revenue streams beyond fixed Power Purchase Agreements (PPAs) by participating actively in energy and ancillary service markets. Technological and power electronics advancements enable wind farms to offer essential grid services and participate as market price-takers \cite{loutan2020avangrid,holttinen2021design}.

Control and operations are crucial for OWF market participation. \cite{kolle2022farmconners} emphasizes farm-level control for profit maximization in FarmConners market analysis. Kim et al. \cite{kim2023economic} highlight the importance of hybrid renewable systems with storage for ancillary market participation, offering new revenue streams. U.S. ancillary markets present significant value opportunities. European wind energy system integration and market participation are discussed by Eguinoa et al. \cite{eguinoa2021wind}, stressing the need for reliable and profitable market engagement through strategic planning and operation controls.

\subsection{Related Work}
Optimization-based models for onshore wind farms with energy storage for participation in multiple energy markets have been extensively studied \cite{dicorato2012planning,nguyen2012new,moghaddam2018optimal,das2020optimal}. Notably, Hou et al. \cite{hou2018cooperation} were among the first to explore OWF market participation with energy storage, though they focused on operational optimization without considering energy storage sizing. Recently, Bechlenberg et al. \cite{bechlenberg2024renewable} addressed energy storage sizing for OWFs optimized for multiple electricity markets. Their work highlights the benefits of integrating OWFs with energy storage, employing a model predictive control approach to manage storage levels and correct prediction uncertainties. However, their ad-hoc approach to electric component sizing, like cable capacity, could lead to sub-optimal system performance and excessive costs.

Varotto et al. \cite{varotto2024optimal} proposed an optimization strategy to evaluate economic benefits through optimal sizing and energy management of Battery Energy Storage Systems (BESS) for hybrid wind-solar offshore farms. A subsequent study \cite{kazemi2025multiobjective} framed a multi-objective optimization problem to assess different storage technologies and locations, yet fixed storage sizes and overlooked control tuning for market participation.

Frequency support for the power grid from renewable sources such as OWFs is increasingly vital. This support can be delivered through wind turbines and integrated sub-stations (VSC/MMC-HVDC) with proper control designs \cite{khan2021analytical,lin2021overview}. We review control schemes and combinations extensively studied in the literature, particularly critiquing those in \cite{khan2021analytical} for fixed sizing assumptions. Through analysis, we identified optimizing control parameters that could enhance OWFs' energy and frequency support participation alongside energy storage design sizing \cite{Lin2022CoordinatedFC}.

Our study focuses on doubly-fed induction generator (DFIG) wind turbines with deloading control, offering significant kinetic energy storage for frequency support with the right controller design. The deloading controller design aims to reserve margin for grid support when needed. Rotor-Side Control (RSC) and Pitch Angle Control (PAC) control designs for DFIG turbines are discussed in \cite{ramtharan2007frequency,ma2010working,zhang2012coordinated}, with coordinated control strategies detailed in \cite{vidyanandan2012primary,zhang2012coordinated}. RSC is preferred for its response speed, particularly with over-speeding for safe operations. Our work examines the primary frequency support of wind farms with an RSC-Droop control strategy.

\textbf{Contributions:} Our literature review identifies two study groups: one focusing on design sizing and market participation but neglecting control parameter optimization, and the other emphasizing control design with fixed system design. We advocate for simultaneous optimization of design sizing and control to improve OWFs' performance as energy market participants. This approach has shown promise in various engineering fields \cite{garciasanz2019} such as offshore wind turbines \cite{abbas2024control}, wave energy converters \cite{strofer2023control}, Energy-Harvesting Ocean Kite \cite{naik2021fused}, etc. Control co-design offers optimal performance in hybrid renewable systems \cite{garcia2024hybrid}, and new market participation formulations for hybrid systems have been proposed \cite{zhu2024enhancing,gao2022multiscale}. These works highlight the importance of market participation in complete assessment and planning, as emphasized by Eguinoa et al. \cite{eguinoa2021wind}. Our work proposes a control co-design approach for optimizing OWF design and control parameters, focusing on market performance.

We extend the multi-timescale formulation from earlier studies \cite{dowling2017multi,sorourifar2018integrated,gao2022multiscale} to a multi-timescale, multi-stage stochastic control co-design formulation, accounting for offshore wind and energy price uncertainty. Addressing uncertainties is crucial for variable energy resources like OWFs. We introduce an optimization formulation to evaluate market participation under uncertainty, considering cable size and onshore energy storage capacity as primary design variables. We also optimize the turbine rotor speed droop-gain parameter and use a droop-based controller to integrate energy storage in reserve markets. This study fills a gap by integrating design and operational planning for OWFs with energy storage, assessing developer revenue. Additionally, we include a post-adhoc validation step for optimized droop gain control. The OWF model and AC-Grid Mini WECC models were developed in a transient EMT-based simulation platform, PSSE. A contingency scenario on the AC grid was simulated to compare estimated reserves against PSSE simulations with optimized droop parameters.

The paper is organized as follows: Section \ref{sec:probdef} defines the problem statement and details the OWF case study. Section \ref{sec:method} outlines the mathematical models and control co-design optimization problem formulation for each step of the multiscale simulation framework. Section \ref{sec:result} presents results, analysis, and validation against the baseline control co-design solutions under different inflation rate considerations. It highlights the role of energy storage sizing and operations. Section \ref{sec:conc} summarizes findings, limitations, and future opportunities.

%% Use \subsection commands to start a subsection.
\section{Problem Definition}\label{sec:probdef}
\begin{figure}[ht]
\centering
\includegraphics[width=0.7\linewidth]{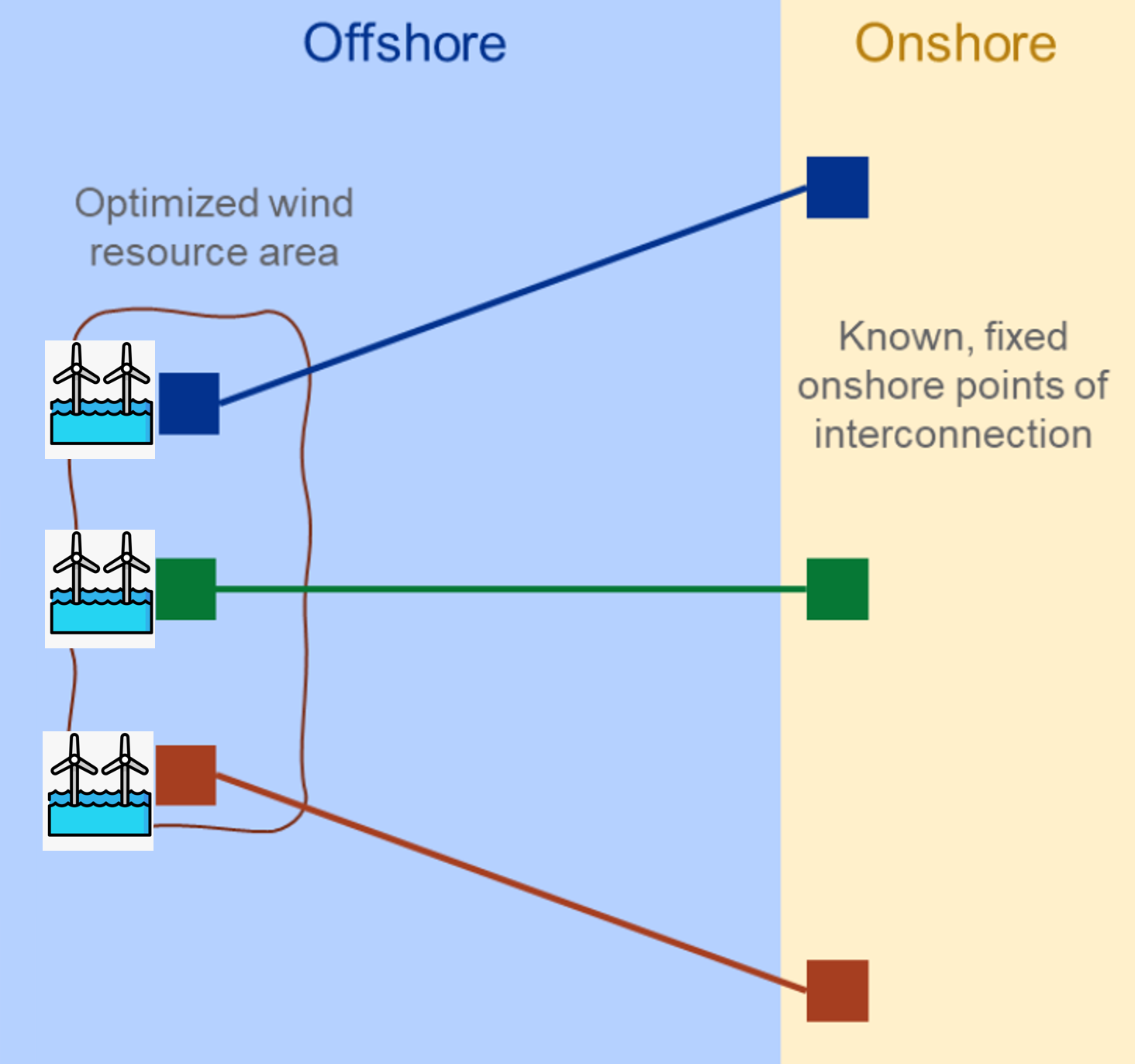}
\caption{The structure of the system considered in this paper. Offshore wind farms are connected to onshore points through HVDC separately.}\label{fig_OWF}
\end{figure}

We examine the control co-design of a proposed OWF on the U.S. West Coast, connected to an onshore point of interconnection (POI) via an HVDC cable configured in a radial topology. The system's topology is illustrated in Figure\ref{fig_OWF}. The choice of the U.S. West Coast use case is inspired by the research conducted by Douville et al. \cite{douville2023offshore}. We consider the installation of an ESS onshore to support the OWF's operations and facilitate energy exchange with the grid \cite{harris2020electricity}. The system, comprising OWF and ESS, participates in the California Independent System Operator (CAISO) electricity markets, including day-ahead (DA) and real-time (RT) energy markets, as well as ancillary service markets by providing upward and downward reserves. For brevity, detailed discussions of these markets are omitted, but further information can be found in Harris et al. \cite{harris2020electricity}. Table \ref{tab_OWFCor} presents the coordinates of the POIs and OWFs, the distances between them, and the rated power of the OWFs, based on the use cases proposed by Douville et al. \cite{douville2023offshore}.

\begin{table}[htbp]
  \centering
  \caption{Names and Coordinates of POIs and OWFs}
  \resizebox{1\textwidth}{!}{
    \begin{tabular}{ccccccc}
    \toprule
    \multirow{2}[4]{*}{POI Name} & \multicolumn{2}{c}{OWF Coordinates} & \multicolumn{2}{c}{POI Coordinates} & \multirow{2}[4]{*}{Distance (km)} & OWF Rated \\
\cmidrule{2-5}          & Latitude & Longitude & Latitude & Longitude &       & Power (MW) \\
    \midrule
    WCASCADE, WA & 42.781 & -124.685 & 47.347 & -122.124 & 545.060 & 1500 \\
    JOHN DAY, OR & 42.836 & -124.660 & 45.678 & -120.738 & 444.423 & 2350 \\
    COTTONWOOD, CA & 42.705 & -124.609 & 40.399 & -122.265 & 324.193 & 1810 \\
    TESLA, CA & 42.630 & -124.558 & 37.712 & -121.565 & 603.598 & 2640 \\
    MOSSLAND, CA & 42.466 & -124.510 & 36.903 & -121.807 & 660.732 & 1800 \\
    \bottomrule
    \end{tabular}}%
  \label{tab_OWFCor}%
\end{table}%

The physical design of the system emphasizes the sizing of the ESS and the HVDC export cables (interties). The primary objective of the control co-design is to minimize installation and material costs while maximizing revenue from energy and ancillary service markets throughout the ESS's operational lifetime. System operations consider power transmission through the HVDC cable and ESS charging/discharging capabilities. Control parameters, specifically the droop control parameters of the OWF and ESS, ensure stable and reliable power and service delivery. In our approach, we propose that both the OWF and ESS jointly provide the reserve. Furthermore, we assume that when the ESS is utilized for reserve provision, it only needs to do so for brief periods, therefore minimally affecting the ESS's state of charge (SoC).

To tackle the challenge of uncertainty in decision-making, this problem is structured as a multi-stage stochastic optimization aimed at minimizing the system's expected net cost. The scenario tree, based on forecast or historical day-ahead (DA), real-time (RT), and up/down reserve prices, in addition to wind speed data, will be thoroughly discussed in Section~\ref{ssec_ScenarioGeneration}.

\section{Methodology}\label{sec:method}
\begin{figure}[h]
    \centering
    \includegraphics[keepaspectratio,width=0.95\linewidth]{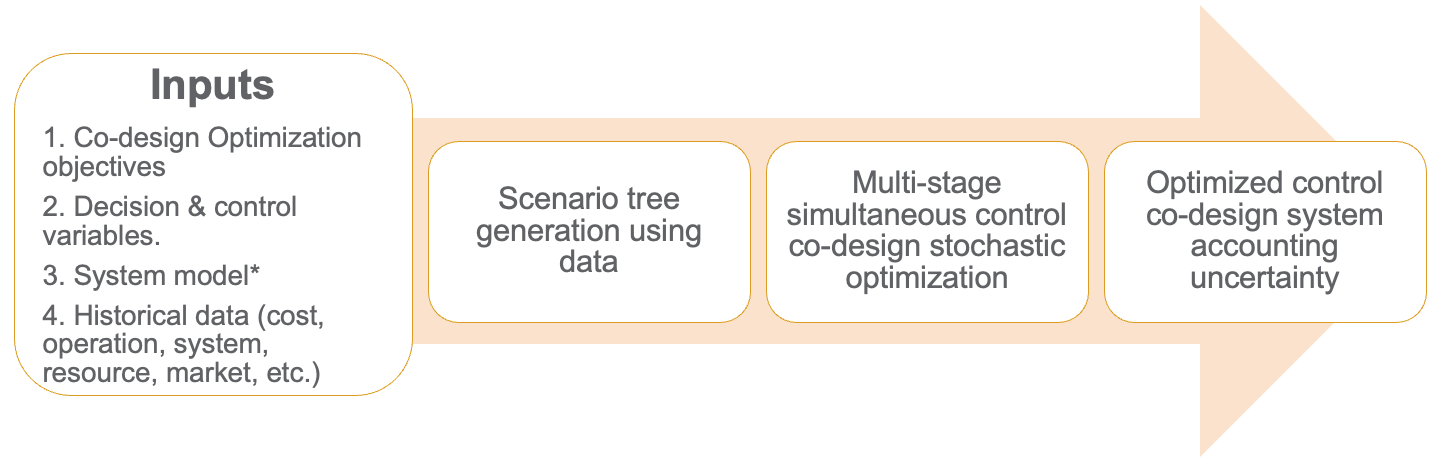}
    \caption{Schematic of the workflow for the proposed multi-time multi-stage stochastic control co-design methodology}
    \label{fig:co-desingOverview}
\end{figure}

The schematic of the proposed methodology is shown in Fig~\ref{fig:co-desingOverview}. The methodology and framework are developed generically such that they can be applied for the simultaneous design and operational control of various energy systems, considering market participation. In this paper the proposed methodology will be described for the control co-design for energy market participation of the OWF with ESS. Next, we will present and discuss the details of individual steps described trough the schematic diagram.

\subsection{Design and control decision variables}
\label{ssec_DesCon}
%  discuss the desing and the control elements.
% \HS{WW: to discuss battery design details. This can focus on the technologies available for storage and the chosen battery technology  }
% \HS{Discuss the cable sizing...}
\textit{Design Variables:} The increasing penetration of renewable energy introduces variability and uncertainty, posing stability challenges for power system operations. ESS are effective in addressing these challenges and have been extensively studied. Proper sizing and optimal operation strategies of ESS integrated with renewable energy systems have shown promising results in enhancing the stability, controllability, and reliability of power systems. Various ESS technologies for wind power integration exist, including pumped hydro storage, compressed air energy storage, flywheel energy storage, and BESS.

In recent years, BESS has experienced significant cost reductions and performance improvements. Due to rapid response time, high efficiency, substantial power and energy density, and convenient siting capability, BESS is widely used for renewable energy integration. Therefore, this study focuses on the design and sizing of lithium-ion batteries as onshore integrated energy storage technology for OWFs. We utilize the PNNL-curated energy storage database \cite{spitsen2020energy} to provide cost and performance estimates for battery storage technologies.

OWFs are large-capacity energy resources requiring efficient power transfer to onshore facilities. Export cable design and sizing are crucial for long-term reliable operations. OWFs, connected to onshore POIs via export cables, can benefit from interconnection or inter-linked export cable topology, as discussed in \cite{douville2023offshore,brinkman2024atlantic}. These connections enable participation in multiple energy markets or facilitate power wheeling \cite{heeter2016wheeling} to avoid power curtailments. Consequently, sizing the export cables solely based on OWF capacity may not be optimal. In this study, cable sizing is considered another design decision variable accounting for market participation. For simplicity, we consider a radial topology connecting OWFs to onshore through subsea export cables, though the framework can accommodate other connection topologies with minor modifications. The optimization of export cable routing and environmental or geographic conditions are not included in this formulation. Cable capacity size ranges and associated costs for the optimization problem were identified using data from \cite{fuchs2024cost,douville2023offshore,brinkman2024atlantic}. The study does not consider design sizing for onshore and offshore converter-inverter substations and other devices; costs associated with these systems are held constant.

\begin{figure}[htbp]
 \centering
   \includegraphics[width=0.8\linewidth]{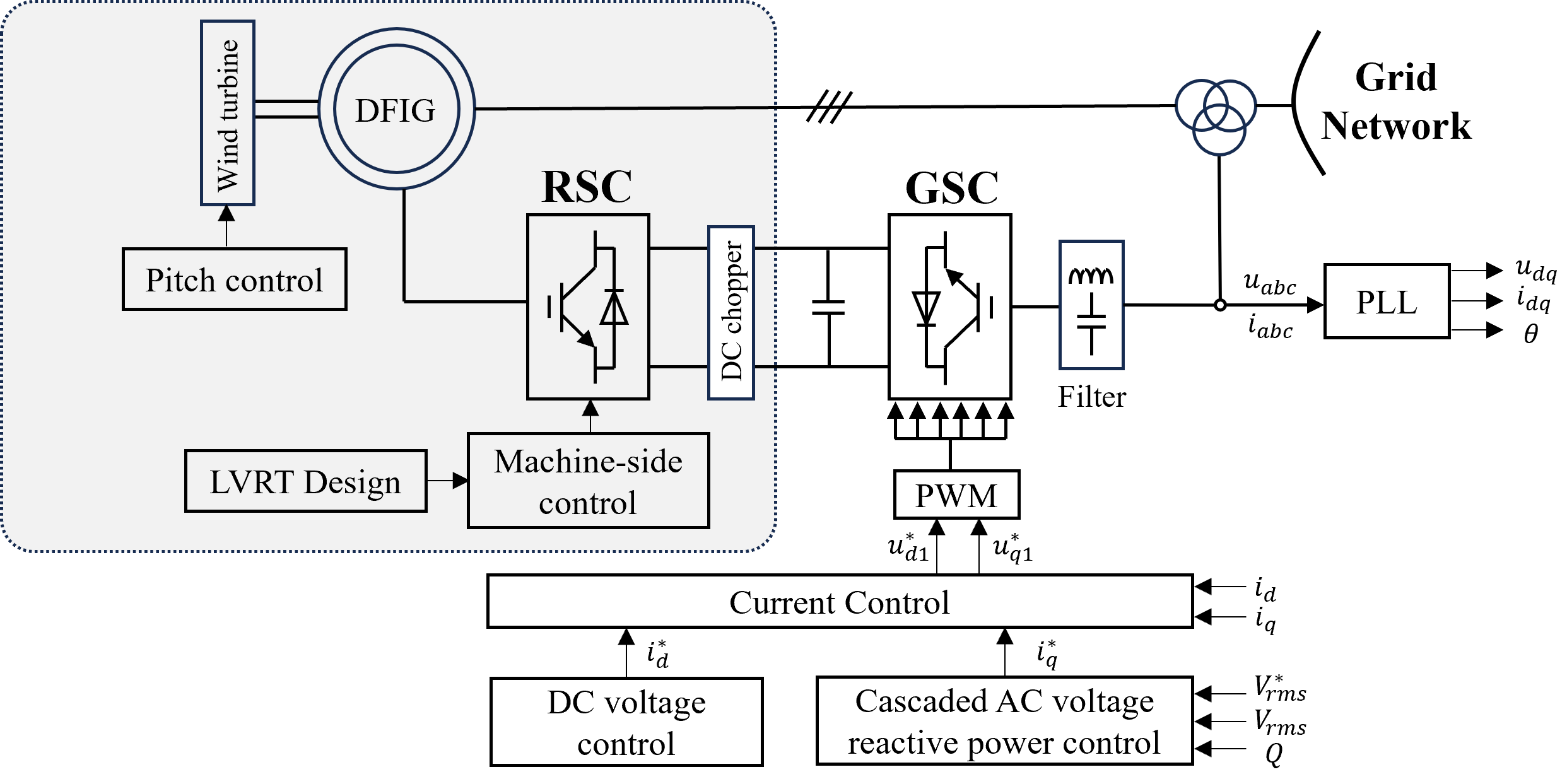}
 \caption{The control diagram of offshore wind turbine is shown highlighting the rotor stator control (RSC). The droop gain ($k$) associate the RSC controller is optimized to provide the required reserve for primary frequency market participation.}
 \label{fig:turbine}
\end{figure}

\textit{Control Variables:} We employ DFIG turbines with partial scale power electronic converters in the rotor circuit while the stator is directly connected to the grid. DFIG-based turbines offer advantages over fixed generator technologies, including speed control within limited ranges, improved energy conversion efficiency, and active/reactive power capabilities. The control system for DFIG-based offshore wind turbines, depicted in Fig.~\ref{fig:turbine}, includes two components: Grid-Side Control (GSC) and Rotor-Side Control (RSC). The GSC maintains the DC voltage across converters by adjusting active power dispatched to the grid and regulates AC voltage through reactive power control. Conversely, the RSC maintains rotor-side AC voltage and optimizes power extraction from wind conditions.

To enhance primary frequency support, the RSC employs a de-loading strategy, establishing a supplementary power reference in response to grid frequency variations. This approach is formalized by the droop relationship outlined in Equation (\ref{eq:droop}), which allocates a power reserve $\Delta P$ from the maximum power point to counter frequency deviations on the grid, in accordance with the guidelines specified in \cite{vidyanandan2012primary}.

% \begin{equation}\label{eq:droop}
% \left\{\begin{array}{l}
%     \Delta P = k \Delta f \\
%     % k =\frac{P^{OWF}}{R}
%     k =P^{OWF}/R
% \end{array}\right.
% \end{equation}

\begin{equation}\label{eq:droop}
\Delta P = k \Delta f = \frac{1}{R} \Delta f
\end{equation}
where $R$ denotes the speed droop parameter to share common load change among multiple generator units.

Despite implementing droop control in the RSC, offshore wind turbines cannot provide frequency support under low or no wind conditions. To overcome this limitation, each OWF is paired with a Grid-Forming Lithium-ion BESS and equipped with a primary droop controller to ensure continuous frequency support capability \cite{zhao2020distributed, mahmood2017decentralized}. The droop controller follows the same frequency support logic as outlined in \eqref{eq:droop}, with its droop gain $k$ jointly optimized alongside the RSC's droop gain.

\subsection{Energy Market : Day Ahead, Real-time and Reserve}
\label{ssec_markets}
The system participates in DA, RT, and ancillary service markets. Before each day begins, operators decide the power allocation for the DA market based on its price, without prior knowledge of RT, up/down reserve prices, and wind speed. Throughout the day, power and reserve allocations for the remaining markets are adjusted based on observed prices and weather conditions at each time step. In this study, we assume a one-hour interval for the DA market and a 15-minute interval for the other markets. The OWF allocates most of its power to the DA market, while the RT market is used primarily to mitigate mismatches between DA market participation and actual power generation. The formulation doesn't model explicit penalties for under performance in the day ahead market, however it is implicitly observed through the change in revenue. 

\begin{figure}[ht]
\centering
\includegraphics[width=0.7\linewidth]{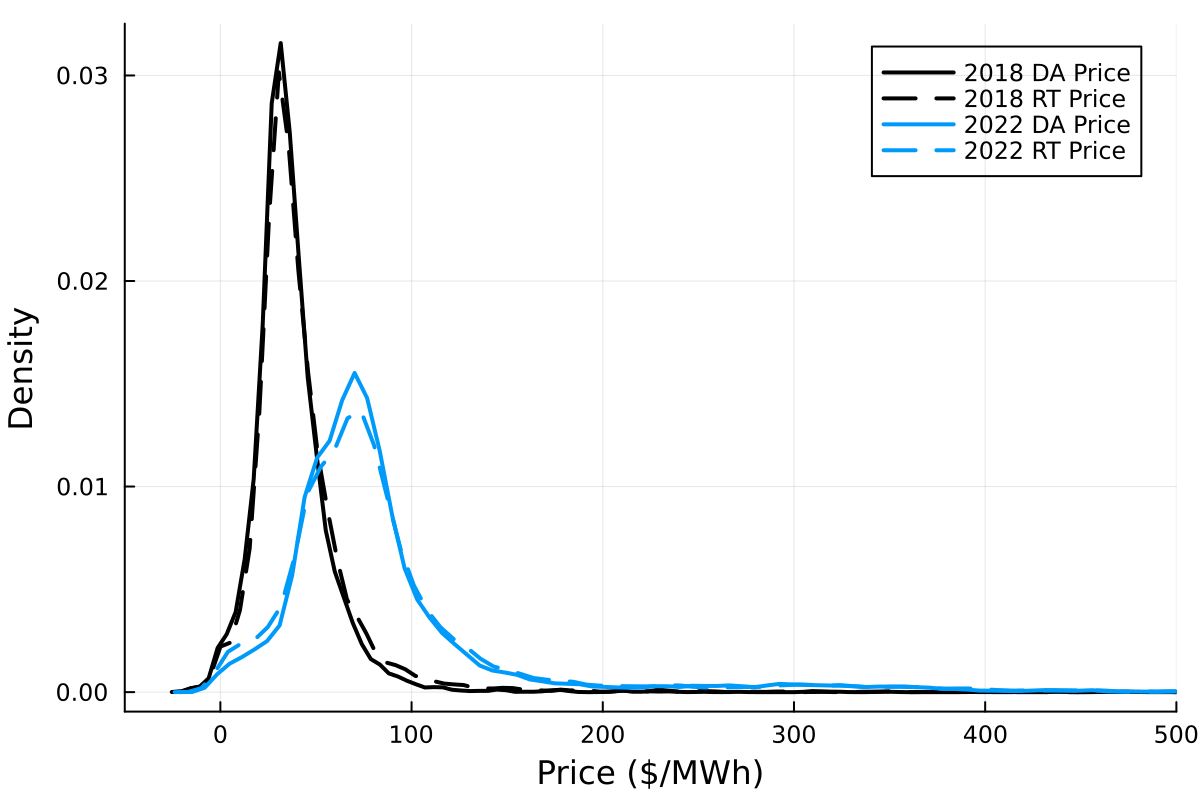}
\caption{Probability density functions of CAISO NP15 region DA and RT market prices for 2018 and 2022.}\label{fig_1822LMP}
\end{figure}

Predicting market power prices and wind speeds is beyond the scope of this paper; hence, we utilize historical data for simulations. We retrieved CAISO price data for the NP15 region for the years 2018 and 2022 from the CAISO OASIS Portal. The CAISO's OASIS API offers real-time and historical data on Locational Marginal Prices (LMPs) for both DA and RT markets. This data includes various market metrics such as energy prices, congestion costs, and losses, updated frequently, with some available on a sub-hourly basis. The year 2018 represents a typical period before the COVID-19 pandemic, while 2022 is the latest year with both price and wind speed data available. Probability density functions of DA and RT market prices for these two years are illustrated in Figure~\ref{fig_1822LMP}, showing a significant energy price increase in 2022 due to rising natural gas costs \cite{california2022}. The impact of high prices on system design and operation will be discussed in Section~\ref{sec:result}.

\subsection{Uncertainties affecting offshore wind farm}
The development and operation of OWFs encounter significant uncertainties affecting their safety, security, and overall viability, with wind speed and market price being primary factors. These uncertainties directly impact the OWF's output and system revenue, thus influencing the required size of the ESS and its capacity. Additionally, effective management of power injection into the DA market, OWF and ESS droop control parameters, and ESS operation is essential to ensure stable and reliable power supply while maximizing total profits. 

Numerous optimization strategies address the challenges posed by these uncertainties in wind farm planning and operation. Techniques such as stochastic optimization \cite{banzo2010stochastic}, robust optimization \cite{chen2015reducing}, and distributionally robust optimization \cite{alismail2017optimal} have been applied. Considering that the lifespan of a BESS typically exceeds ten years, evaluating the long-term expected revenue of the OWF yields more effective design outcomes. Furthermore, design and market participation decisions are often not made simultaneously. Therefore, we employ a multistage stochastic optimization framework for our problem, which assesses uncertainties at various decision-making stages and optimizes the overall expected performance.

\subsection{Scenario generation scheme}
\label{ssec_ScenarioGeneration}
% \HS{Bowen: To write the scenario generation overview and discuss detail of the choice and the scheme used for the work.}
The scenario generation process is crucial for modeling and optimizing the performance of OWFs. Scenarios represent various potential future states of key variables, such as wind speed and energy market prices, facilitating robust decision-making under uncertainty. This process involves creating a set of representative scenarios that captures the variability and stochastic nature of these variables, enabling effective optimization and control of OWFs.

The scenario generation scheme aims to balance accuracy with computational efficiency. The following steps outline the adopted approach:

\textit{Data Collection and Normalization:} Initially, historical data on wind speeds and market prices is collected. Offshore wind data focuses on heights relevant to turbine operations (e.g., 140 meters). This data is then normalized for consistency across different periods and locations.

\textit{Probability Distribution Identification:} Identifying suitable probability distributions for key variables is crucial. The Weibull distribution is often used for wind speed modeling due to its appropriateness for characterizing wind behavior. The goodness-of-fit is verified using statistical tests like the Kolmogorov-Smirnov test, additionally if necessary alternative distributions can also be considered.

\textit{Scenario Generation Techniques:} Once the distribution is identified and validated, scenarios are generated using sampling methods. Monte Carlo simulation is employed to create a wide range of scenarios by randomly sampling from the identified distribution, providing a comprehensive view of potential future outcomes.

\textit{Scenario Reduction and Clustering:} Due to the large number of generated scenarios, it's necessary to reduce the dataset to a manageable size without losing critical information. This is accomplished through techniques such as clustering, where similar scenarios are grouped, and representative scenarios are selected from each cluster. The goal is to retain the diversity of the original scenarios while enhancing computational efficiency in the optimization process.

The generated scenarios are applied to the optimization of OWF layouts and operations. By simulating different OWF configurations under these scenarios, designs robust to uncertainties in wind speeds and market conditions are identified. This approach ensures that OWFs maximize energy production and market participation while minimizing risks and costs. A simplified block diagram of the scenario generation scheme is presented in Figure~\ref{fig:diagram1}. Detailed algorithmic descriptions of the scenario tree generation are provided in appendix-\ref{apd:scenario} as Algorithm \ref{alg:kde}.
\begin{figure}[ht]
\centering
\includegraphics[width=0.97\linewidth]{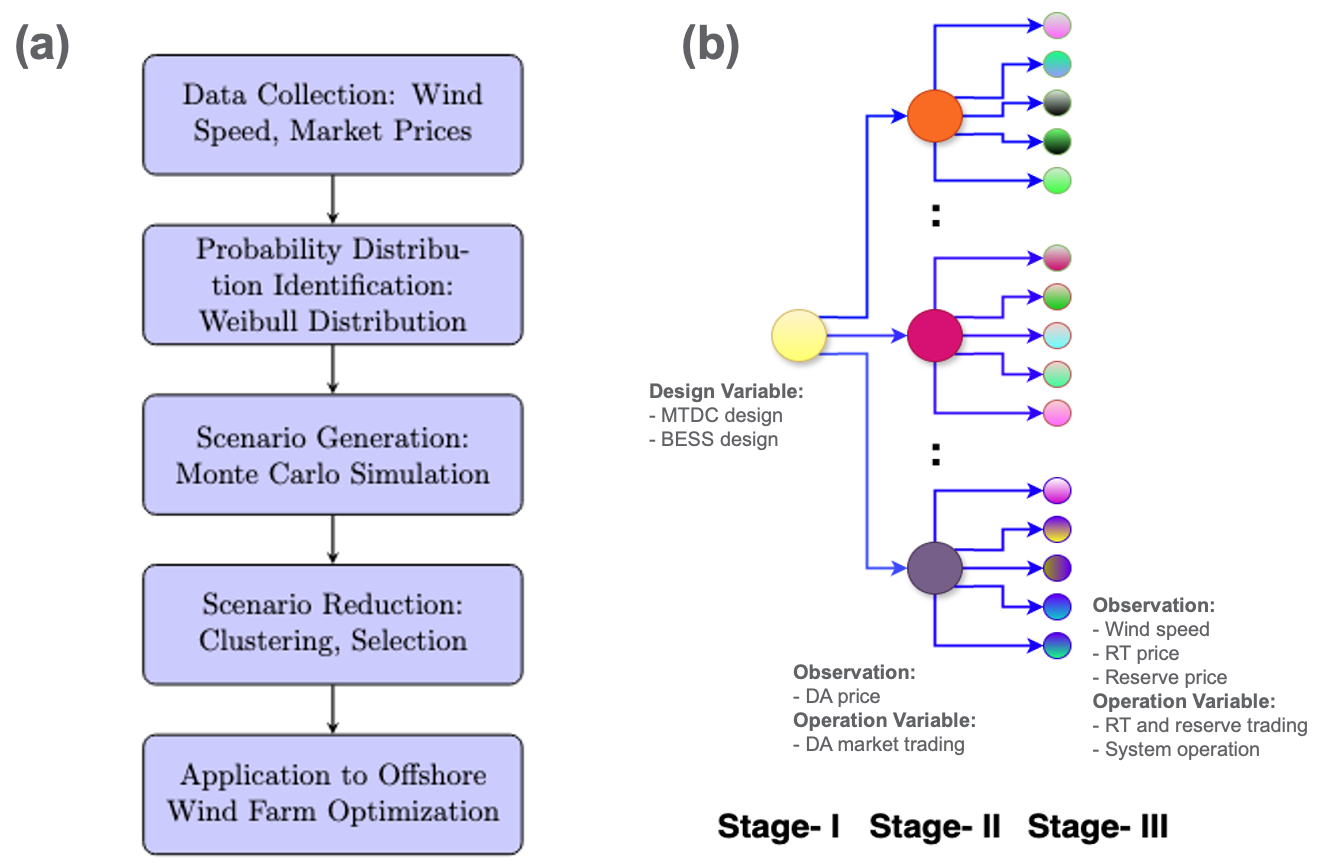}
\caption{(a) Scenario generation scheme for offshore Wind Farm Optimization. (b) Schematic scenario tree with defining the multi-stage decision variables.}\label{fig:diagram1}
\end{figure}

\subsection{Multi-time, Multi-stage stochastic control co-design formulation }
We next discuss the details of proposed control co-design  multi-time scale, multi-stage stochastic formulation.
\subsubsection{Multi-scale formulation}
It is common for different markets to clear at varying frequencies, and wind speed forecast data may not align with these markets. In this paper, we consider DA price and wind speed data are hourly, while RT and up/down reserve prices are provided every 15 minutes. 

Following \cite{dowling2017multi}, we use $\mathcal T_1=\{1,2,\cdots,24\}$ and $\mathcal T_2=\{1,2,\cdots, 96\}$ to represent the index sets of each hour and each 15-minute interval in a day, respectively. Additionally, the set $\mathcal T_{21}(t)=\{4(t-1)+k\mid k=1,2,3,4\}$ contains all the four 15-minute indices in hour $t$. This approach can be easily extended to accommodate more time scales and other time resolutions.

\subsubsection{Multi-stage formulation with random variables and uncertain parameters}
There are three stages in our problem: design, DA, and RT. Random events at DA and RT stages are represented as $\omega^\mathrm{DA}$ and $\omega^\mathrm{RT}$, respectively. When making the design decision, no uncertain parameter can be observed. At the DA stage, DA price $\boldsymbol\lambda^\mathrm{DA}(\omega^\mathrm{DA})$ for the next 24 hours is cleared and known. At RT stage, all other uncertain parameters are revealed, including wind power $\mathbf P^\mathrm{OWF}(\omega^\mathrm{DA},\omega^\mathrm{RT})$ extracted from OWF by maximum power point tracking, RT price $\boldsymbol\lambda^\mathrm{RT}(\omega^\mathrm{DA},\omega^\mathrm{RT})$, up reserve price $\boldsymbol\lambda^\mathrm{ResU}(\omega^\mathrm{DA},\omega^\mathrm{RT})$, and down reserve price $\boldsymbol\lambda^\mathrm{ResD}(\omega^\mathrm{DA},\omega^\mathrm{RT})$.
\subsubsection{Control co-design formulation}
In this section, we provide the detailed formulation of the problem. Let $\mathbb E_{\omega^\cdot}(X)$ denote the expectation of $X$ over all scenarios of $\omega^\cdot$. The model is as follows
\begin{align}
\min\quad&R^\mathrm{tax}\left(\lambda^\mathrm{E}sz^\mathrm{E}+\lambda^\mathrm{C}sz^\mathrm{C}\right)-R^\mathrm{Year}\mathbb E_{\omega^\mathrm{DA}}\left[\mathcal F\left(sz^\mathrm{E},sz^\mathrm{C};\mathrm\omega^\mathrm{DA}\right)\right]\label{obj_1}\\
\mathrm{s.t.}\quad&0\leq sz^\mathrm{E}\leq 0.05P^\mathrm{WFR},\ sz^\mathrm{C}\geq 0\label{size_lim}\\
&(\ref{end_soc}-\ref{cycle_limit})\nonumber
\end{align}
In the objective function (\ref{obj_1}), the first term is the installation cost of ESS and HVDC transmission cable, while the second term is the present value of the expected revenue from the OWF over the lifetime of the ESS. Specifically, $R^\mathrm{tax}$ is a tax credit coefficient provided for new, qualified clean energy properties. Coefficients $\lambda^\mathrm E$ and $\lambda^\mathrm C$ represent the unit costs of ESS and cable, respectively. Variables $sz^\mathrm E$ and $sz^\mathrm C$ denote the rated power of ESS and HVDC cable, respectively. $R^\mathrm{Year}$ is an annuity factor that converts expected daily revenue to present value over the lifetime of the ESS. If it has a lifetime of $Y$ years and the inflation rate is $r$, then $R^\mathrm{Year}=365\frac{(1+r)^Y-1}{r(1+r)^Y}$. The constraints (\ref{size_lim}) limit the ESS to below 5\% of OWF rated power $P^\mathrm{WFR}$. The limit allows us to have sufficient battery size exploration, while considering the cost of large batteries. Constraints (\ref{end_soc}-\ref{cycle_limit}) are about ESS expected SoC at the end of day and expected operation cycles, which will be discussed later.

For given rated power of ESS and HVDC cable $(sz^\mathrm{E},sz^\mathrm{C})$ and a realization of DA scenario $\omega^\mathrm{DA}$, $\mathcal F\left(sz^\mathrm{E},sz^\mathrm{C};\mathrm\omega^\mathrm{DA}\right)$ calculates the daily revenue of the OWF as below.
\begin{align}
\mathcal F\left(sz^\mathrm{E},sz^\mathrm{C};\omega^\mathrm{DA}\right)=\max\ &\sum_{t\in\mathcal T_1}\lambda^\mathrm{DA}_{t}\left(\omega^\mathrm{DA}\right)p_{t}^\mathrm{WD}\nonumber\\
&+\mathbb E_{\omega^\mathrm{RT}\mid\omega^\mathrm{DA}}\left[\mathcal G\left(sz^\mathrm{E},sz^\mathrm{C},\mathbf p^\mathrm{WD};\omega^\mathrm{DA},\omega^\mathrm{RT}\right)\right]\label{obj_2}
\end{align}
The first and second terms are DA and expected RT stage revenues, respectively. Variable $p_{t}^\mathrm{WD}$ represents the power traded to the DA market at time $t$.

For given rated power of ESS and HVDC cable $(sz^\mathrm{E},sz^\mathrm{C})$, power traded to DA market $\mathbf p^\mathrm{WD}$, and realizations of DA scenario $\omega^\mathrm{DA}$ and RT scenario $\omega^\mathrm{RT}$, $\mathcal G(sz^\mathrm{E},\allowbreak sz^\mathrm{C},\allowbreak \mathbf p^\mathrm{WD};\allowbreak \omega^\mathrm{DA},\allowbreak \omega^\mathrm{RT})$ calculates the RT stage revenue of the OWF. Its objective functions is
\begin{align}
\max\quad&\sum_{t\in\mathcal T_2}\lambda^\mathrm{RT}_{t}\left(\omega^\mathrm{DA},\omega^\mathrm{RT}\right)p_{t}^\mathrm{WR}+\sum_{t\in\mathcal T_2}\lambda^\mathrm{ResU}_{t}\left(\omega^\mathrm{DA},\omega^\mathrm{RT}\right)\left(p^\mathrm{ResWU}_{t}+p^\mathrm{ResBU}_{t}\right)\nonumber\\
&+\sum_{t\in\mathcal T_2}\lambda^\mathrm{ResD}_{t}(\omega^\mathrm{DA},\omega^\mathrm{RT})(p^\mathrm{ResWD}_{t}+p^\mathrm{ResBD}_{t})
\end{align}
These three terms calculate RT, up reserve, and down reserve revenues, respectively. Variable $p_{t}^\mathrm{WR}$ represents the power traded to the RT market at time $t$; $p^\mathrm{ResWU}_{t}$ and $p^\mathrm{ResBU}_{t}$ denote the up reserve provided by OWF and ESS, respectively; and $p^\mathrm{ResWD}_{t}$, and $p^\mathrm{ResBD}_{t}$ are the down reserve provided by them.

Problem $\mathcal G(sz^\mathrm{E},\allowbreak sz^\mathrm{C},\allowbreak \mathbf p^\mathrm{WD};\allowbreak \omega^\mathrm{DA},\allowbreak \omega^\mathrm{RT})$ has a series of constraints. The power at the onshore and offshore sides should be balanced and transmitted through the HVDC cable.
\begin{align}
&p^\mathrm{W}_{t_1}=V_{1,t_2}(V_{1,t_2}-V_{2,t_2})g\ \forall t_1\in\mathcal T_1,\ t_2\in\mathcal T_{21}(t_1)\label{on_bal}\\
&(p^\mathrm{dis}_{t_2}-p^\mathrm{ch}_{t_2})-(p^\mathrm{WR}_{t_2}+p^\mathrm{WD}_{t_1})=V_{2,t_2}(V_{2,t_2}-V_{1,t_2})g\ \forall t_1\in\mathcal T_1,\ t_2\in\mathcal T_{21}(t_1)\label{off_bal}\\
&1.1\ (V_{1,t}(V_{1,t}-V_{2,t})g+p^\mathrm{ResWU}_{t})\leq sz^\mathrm C\ \forall t\in\mathcal T_2\label{cal_lim}\\
&\underline V_i\leq V_{i,t}\leq\overline V_i\ \forall i\in\{1,2\}
,\ t\in\mathcal T_2\label{vol_lim}
\end{align}
In these constraints, parameter $g$ is the conductance of the transmission cable; $\underline V_i$ and $\overline V_i$ are lower and upper voltage limits of bus $i$, respectively. Here $i=1$ refers to the offshore bus and $i=2$ refers to the onshore bus. At each time $t$, variable $p^\mathrm{W}_{t}$ represents the power actually generated by the OWF; $V_{i,t}$ is the voltage of node $i$; $p^\mathrm{dis}_{t}$ and $p^\mathrm{ch}_{t}$ denote the discharged and charged power of ESS, respectively.

Constraints (\ref{on_bal}) and (\ref{off_bal}) ensure the power balance on the offshore and onshore sides, respectively. Constraint (\ref{cal_lim}) limits the power transmitted through the HVDC cable, accounting for a 10\% as additional factor of safety. Voltage limits are imposed by constraint (\ref{vol_lim}).

As stated in Section~\ref{ssec_markets}, RT market is primarily used to mitigating the difference between DA market participation and actual OWF generation. To prevent excessive power trading in the RT market to exploit the price differences between these two markets, the following limits are imposed.
\begin{align}
&p_{t_2}^\mathrm{WR}\leq\max\{P^\mathrm{OWF}_{t_1}(\omega^\mathrm{DA},\omega^\mathrm{RT})-P^\mathrm{OWF}_{t_1}(\omega^\mathrm{DA}),0\}\ \forall t_1\in\mathcal T_1,\ t_2\in\mathcal T_{21}(t_1)\\
&p_{t_2}^\mathrm{WR}\geq\min\{P^\mathrm{OWF}_{t_1}(\omega^\mathrm{DA},\omega^\mathrm{RT})-P^\mathrm{OWF}_{t_1}(\omega^\mathrm{DA}),0\}\ \forall t_1\in\mathcal T_1,\ t_2\in\mathcal T_{21}(t_1)
\end{align}
Here $P^\mathrm{OWF}_{t}(\omega^\mathrm{DA})=\mathbb E_{\omega^\mathrm{RT}}[P^\mathrm{OWF}_{t}(\omega^\mathrm{DA},\omega^\mathrm{RT})]$. These two constraints ensure that the power traded in the RT market remains within the difference between the actual and expected wind power in each DA scenario.

Considering up and down reserves, the OWF's actual output should stay within the following range.
\begin{align}
&p_{t_1}^\mathrm W+p^\mathrm{ResWU}_{t_2}\leq P^\mathrm{OWF}_{t_1}(\omega^\mathrm{DA},\omega^\mathrm{RT})\ \forall t_1\in\mathcal T_1,\ t_2\in\mathcal T_{21}(t_1)\\
&p_{t_1}^\mathrm W-p^\mathrm{ResWD}_{t_2}\geq 0\ \forall t_1\in\mathcal T_1,\ t_2\in\mathcal T_{21}(t_1)
\end{align}

As discussed in Section~\ref{ssec_DesCon}, we have the following droop relationship when the OWF and ESS provide fast frequency support.
\begin{align}
&p^\mathrm{ResWU}_{t}=k^\mathrm{ResW}_t\Delta f^U_{\max}\ \forall t\in\mathcal T_2\label{WU_Res}\\
&p^\mathrm{ResWD}_{t}=k^\mathrm{ResW}_t\Delta f^D_{\max}\ \forall t\in\mathcal T_2\label{WD_Res}\\
&p^\mathrm{ResBU}_{t}=k^\mathrm{ResB}_t\Delta f^U_{\max}\ \forall t\in\mathcal T_2\label{BU_Res}\\
&p^\mathrm{ResBD}_{t}=k^\mathrm{ResB}_t\Delta f^D_{\max}\ \forall t\in\mathcal T_2\label{BD_Res}\\
&k^\mathrm{ResW}_{t_2}\geq P^\mathrm{OWF}_{t_1}(\omega^\mathrm{DA},\omega^\mathrm{RT})/R_{max}\ \forall t_1\in\mathcal T_1,\ t_2\in\mathcal T_{21}(t_1)\label{kWL}\\
&k^\mathrm{ResW}_{t_2}\leq P^\mathrm{OWF}_{t_1}(\omega^\mathrm{DA},\omega^\mathrm{RT})/R_{min}\ \forall t_1\in\mathcal T_1,\ t_2\in\mathcal T_{21}(t_1)\label{kWD}\\
&sz^\mathrm E/R_{max}\leq k^\mathrm{ResB}_t\leq sz^\mathrm E/R_{min}\ \forall t\in\mathcal T_2\label{kBLim}\\
&k^\mathrm{ResW}_{t_2}+k^\mathrm{ResB}_{t_2}\geq P^\mathrm{OWF}_{t_1}(\omega^\mathrm{DA},\omega^\mathrm{RT})/R_{all}\ \forall t_1\in\mathcal T_1,\ t_2\in\mathcal T_{21}(t_1)\label{kTolLim}
\end{align}
Parameters $\Delta f^U_{\max}$ and $\Delta f^D_{\max}$ represent maximum up and down frequency deviations under planned disturbances, respectively. Variables $k^\mathrm{ResW}_t$ and $k^\mathrm{ResB}_t$ are the droop control parameters of OWF and ESS, respectively. Constraints (\ref{WU_Res}-\ref{BD_Res}) allocate up and down reserves to the OWF and ESS based on droop control parameters and maximum frequency deviations. Constraints (\ref{kWL}-\ref{kTolLim}) limit the percentage of OWF and ESS primary frequency support. Additionally, constraint \eqref{kTolLim} ensures that each OWF, when combined with its ESS, provides joint mandatory primary frequency support to the main grid.

The ESS has the following dynamic and state constraints.
\begin{align}
&SoC_{t}-SoC_{t-1}=(\eta^\mathrm{ch}p_{t}^\mathrm{ch}-\eta^\mathrm{dis}p_{t}^\mathrm{dis})T^\mathrm{length}_2\ \forall t\in\mathcal T_2\backslash\{1\}\label{batt_dyn}\\
&SoC_{1}-0.5sz^\mathrm E\mathrm{DurH}=(\eta^\mathrm{ch}p_{1}^\mathrm{ch}-\eta^\mathrm{dis}p_{1}^\mathrm{dis})T^\mathrm{length}_2\label{batt_dyn1}\\
&p_{t}^\mathrm{ch}+p^\mathrm{ResBD}_{t}\leq sz^\mathrm E\quad\forall t\in\mathcal T_2\label{ch_lim}\\
&p_{t}^\mathrm{dis}+p^\mathrm{ResBU}_{t}\leq sz^\mathrm E\quad\forall t\in\mathcal T_2\label{dis_lim}\\
&0\leq SoC_{t}\leq sz^\mathrm E\mathrm{DurH}\quad\forall t\in\mathcal T_2\label{soc_lim}
\end{align}
Parameters $\eta^\mathrm{ch}$ and $\eta^\mathrm{dis}$ represent the charging and discharging efficiency of the ESS, respectively; $\mathrm{DurH}$ denotes the duration hour of the ESS; $T^\mathrm{length}_2$ is the time length of the time indices in set $\mathcal T_2$, which is $1/4$ hour here. Variable $SoC_t$ is the state of charge of ESS. We assume the ESS is initially half charged. Constraints (\ref{batt_dyn}-\ref{batt_dyn1}) are ESS dynamic equations and (\ref{ch_lim}-\ref{soc_lim}) impose charging, discharging, and SoC limits, respectively.

Finally we have the following two constraints. Constraint (\ref{end_soc}) requires the expected SoC of ESS to be half of its energy capacity at the end of the day and constraint (\ref{cycle_limit}) ensures its expected total discharge power does not exceed daily cycle limit $\mathrm{cyl}^\mathrm{lim}$.
\begin{align}
&\mathbb E_{\omega^\mathrm{RT}}[SoC_{|\mathcal T_2|}]=0.5sz^\mathrm E\mathrm{DurH}\label{end_soc}\\
&\mathbb E_{\omega^\mathrm{RT}}\left[\sum_{t\in\mathcal T_2}p^\mathrm{dis}_t\eta^\mathrm{dis}T^\mathrm{length}_2\right]\leq sz^\mathrm E\mathrm{cyl}^\mathrm{lim}\label{cycle_limit}
\end{align}

\section{Results}\label{sec:result}
\subsection{Baseline, CCD, and comparison cases setting}
% \HS{HS, WW: To complete the section, discuss the benefits and comparisons}
To illustrate the advantages of utilizing the Control Co-Design (CCD) model discussed in the previous section and to explore the benefits of pairing the ESS with the OWF, we computed and compared results for the following cases. The droop parameters are set as: $R_{min}=0.1$ and $R_{max}=0.5$ for wind farms; $R_{min} = 0.01$ and $R_{max}=0.5$ for ESS; $R_{all}=0.2$. We generated 20 DA stage scenarios and, for each DA scenario, 5 RT stage scenarios, resulting in a total of 100 scenarios.
\begin{itemize}
    \item Base: Setting ESS rated power $sz^\mathrm E$ to be 2\% of OWF rated power; fixing $k^\mathrm{ResB}$ and $k^\mathrm{ResW}$ at their upper limits for providing maximum frequency support; using one or more HVDC cables with fixed rated power of 2600 MW; 3\% inflation; wind and price data of 2018.
    \item CCD18-3, CCD18-5, and CCD18-8: CCD model with 3\%, 5\%, and 8\% reflation rate \footnote{Reflation refers to a period of economic recovery where monetary and fiscal policies are used to stimulate growth and combat deflation, often characterized by increasing prices.}; wind and price data of 2018.
    \item CCD22-3, CCD22-5, and CCD22-8: CCD model with 3\%, 5\%, and 8\% reflation rate; wind and price data of 2022.
    \item No reserve: Same as CCD18-3 case, but no reserve is provided for frequency support.
    \item No ESS: Same as CCD18-3 case, but no ESS installed.
\end{itemize}

\subsection{Design results and discussion}
The design results for the ESS and the net profits of OWFs are detailed in Table~\ref{tab_BESS}, while the cable design results are shown in Table~\ref{tab_Cable}. A total cost of \$855.4 million for both onshore and offshore converters is included in all cases.

Examining the first four cases in Table~\ref{tab_BESS}, the base case, which provides the largest reserve for frequency support, yields the lowest profit. Conversely, the no reserve case, which allocates all power to market sales, achieves the highest profit. The no ESS case generates less profit than CCD18-3 due to the absence of ESS. Significantly higher profits are observed in cases using 2022 data, reflecting the much higher energy prices compared to 2018.

\begin{table}[htbp]
  \centering
  \caption{ESS Size Solution and Offshore Wind Farm Profit}
  \resizebox{0.8\textwidth}{!}{
    \begin{tabular}{|c|c|c|c|c|}
    \hline
    \multirow{2}{*}{Case} & \multirow{2}[4]{*}{POI} & \multicolumn{2}{c|}{Battery} & \multirow{2}[4]{*}{Net Profit(\$M)} \\
\cline{3-4}          &       & Rated Power(MW) & Cost(\$M) &  \\
    \hline
    \multirow{5}{*}{CCD18-3} & WCASCADE & 45.00 & 40.01 & 3173.67 \\
\cline{2-5}          & John Day & 70.50 & 62.68 & 5266.97 \\
\cline{2-5}          & COTWDPGE & 54.30 & 48.28 & 4090.65 \\
\cline{2-5}          & Tesla & 79.20 & 70.41 & 5654.59 \\
\cline{2-5}          & Mossland & 54.00 & 48.01 & 3248.39 \\
    \hline
    \hline
    \multirow{5}{*}{CCD18-5} & WCASCADE & 45.00 & 40.01 & 2642.18 \\
\cline{2-5}          & John Day & 70.50 & 62.68 & 4472.73 \\
\cline{2-5}          & COTWDPGE & 54.30 & 48.28 & 3460.36 \\
\cline{2-5}          & Tesla & 79.20 & 70.41 & 4787.86 \\
\cline{2-5}          & Mossland & 54.00 & 48.01 & 2689.99 \\
    \hline
    \hline
    \multirow{5}{*}{CCD18-8} & WCASCADE & 40.08 & 35.63 & 2021.44 \\
\cline{2-5}          & John Day & 61.20 & 54.41 & 3545.20 \\
\cline{2-5}          & COTWDPGE & 52.53 & 46.71 & 2724.14 \\
\cline{2-5}          & Tesla & 65.90 & 58.59 & 3776.06 \\
\cline{2-5}          & Mossland & 39.94 & 35.51 & 2038.68 \\
    \hline
    \hline
    \multirow{5}{*}{Base} & WCASCADE & 30.00 & 26.67 & 2887.38 \\
\cline{2-5}          & John Day & 47.00 & 41.79 & 5079.29 \\
\cline{2-5}          & COTWDPGE & 36.20 & 32.18 & 3879.75 \\
\cline{2-5}          & Tesla & 52.80 & 46.94 & 5024.09 \\
\cline{2-5}          & Mossland & 36.00 & 32.01 & 2991.06 \\
    \hline
    \hline
    \multirow{5}{*}{No Reserve} & WCASCADE & 6.12  & 5.44  & 3223.98 \\
\cline{2-5}          & John Day & 6.20  & 5.51  & 5339.90 \\
\cline{2-5}          & COTWDPGE & 14.40 & 12.80 & 4150.18 \\
\cline{2-5}          & Tesla & 5.55  & 4.94  & 5735.24 \\
\cline{2-5}          & Mossland & 5.32  & 4.73  & 3301.59 \\
    \hline
    \hline
    \multirow{5}{*}{No Battery} & WCASCADE & 0.00  & 0.00  & 3142.00 \\
\cline{2-5}          & John Day & 0.00  & 0.00  & 5219.27 \\
\cline{2-5}          & COTWDPGE & 0.00  & 0.00  & 4052.53 \\
\cline{2-5}          & Tesla & 0.00  & 0.00  & 5604.78 \\
\cline{2-5}          & Mossland & 0.00  & 0.00  & 3216.57 \\
    \hline
    \hline
    \multirow{5}{*}{CCD22-3} & WCASCADE & 75.00 & 66.68 & 8755.87 \\
\cline{2-5}          & John Day & 117.50 & 104.46 & 12955.41 \\
\cline{2-5}          & COTWDPGE & 90.50 & 80.46 & 10872.30 \\
\cline{2-5}          & Tesla & 132.00 & 117.36 & 14410.31 \\
\cline{2-5}          & Mossland & 90.00 & 80.02 & 8592.07 \\
    \hline
    \hline
    \multirow{5}{*}{CCD22-5} & WCASCADE & 75.00 & 66.68 & 7548.84 \\
\cline{2-5}          & John Day & 117.50 & 104.46 & 11230.12 \\
\cline{2-5}          & COTWDPGE & 90.50 & 80.46 & 9421.34 \\
\cline{2-5}          & Tesla & 132.00 & 117.36 & 12483.37 \\
\cline{2-5}          & Mossland & 90.00 & 80.02 & 7386.21 \\
    \hline
    \hline
    \multirow{5}{*}{CCD22-8} & WCASCADE & 45.00 & 40.01 & 6140.86 \\
\cline{2-5}          & John Day & 70.50 & 62.68 & 9220.25 \\
\cline{2-5}          & COTWDPGE & 54.30 & 48.28 & 7729.58 \\
\cline{2-5}          & Tesla & 79.20 & 70.41 & 10236.87 \\
\cline{2-5}          & Mossland & 54.00 & 48.01 & 5981.84 \\
    \hline
    \end{tabular}}%
  \label{tab_BESS}%
\end{table}%

\begin{table}[htbp]
  \centering
  \caption{Cable Size Solution}
  \resizebox{1\textwidth}{!}{
    \begin{tabular}{|c|c|c|c|c|}
    \hline
    \multirow{2}[4]{*}{Case} & \multirow{2}[4]{*}{POI} & \multicolumn{3}{c|}{Cable} \\
\cline{3-5}          &       & Capacity(MW) & Material Cost(\$M) & Installation Cost(\$M) \\
    \hline
    \multirow{5}{*}{\makecell{CCD18-*\\CCD22-*\\No Reserve\\No Battery}} & WCASCADE & 1650.00 & 279.35 & 64.39 \\
\cline{2-5}          & John Day & 2585.00 & 356.85 & 52.50 \\
\cline{2-5}          & COTWDPGE & 1991.00 & 200.49 & 38.30 \\
\cline{2-5}          & Tesla & 2904.00 & 544.46 & 71.30 \\
\cline{2-5}          & Mossland & 1980.00 & 406.36 & 78.05 \\
    \hline
    \hline
    \multirow{5}[10]{*}{Base} & WCASCADE & 2600.00 & 440.19 & 64.39 \\
\cline{2-5}          & John Day & 2600.00 & 358.92 & 52.50 \\
\cline{2-5}          & COTWDPGE & 2600.00 & 261.82 & 38.30 \\
\cline{2-5}          & Tesla & 5200.00 & 974.93 & 71.30 \\
\cline{2-5}          & Mossland & 2600.00 & 533.61 & 78.05 \\
    \hline
    \end{tabular}}%
  \label{tab_Cable}%
\end{table}%

Except for the base case, all other cases utilize the smallest transmission cable size that satisfies the minimum requirement specified in constraint~(\ref{cal_lim}), thereby reducing costs. There are two critical values for ESS size: 3\% and the upper limit of 5\% of the OWF rated power. In most scenarios, the ESS size is set at one of these two values.

\subsection{ESS sizing analysis}
The size of the ESS primarily depends on the revenue it can generate during daily operations. On one hand, the system charges the ESS when energy prices are low and discharges it during peak price hours. This revenue strategy relies on the energy price differential across one or more consecutive days. On the other hand, if the ESS provides the reserve for frequency support, the OWF can allocate more power to market sales rather than retaining generation capacity, with this revenue influenced by the absolute energy price.

When energy prices vary within specific ranges, as observed in cases CCD18-3, CCD18-5, and CCD22-8, constraints (\ref{kWL}) and (\ref{kTolLim}) are binding, enabling the OWF to send most of its power to the markets. An ESS sized at 3\% of the OWF's rated power can adequately meet the reserve requirements specified in constraint (\ref{kTolLim}). If the revenue generated by the ESS does not justify its cost, the system opts for a smaller size, as evidenced by case CCD18-8. Conversely, higher revenue potential supports a larger ESS size, even reaching the upper limit, as demonstrated in cases CCD22-3 and CCD22-5.

\subsection{Market participation and ESS operation}
\begin{figure}[ht]
    \centering
    \includegraphics[width=0.55\linewidth]{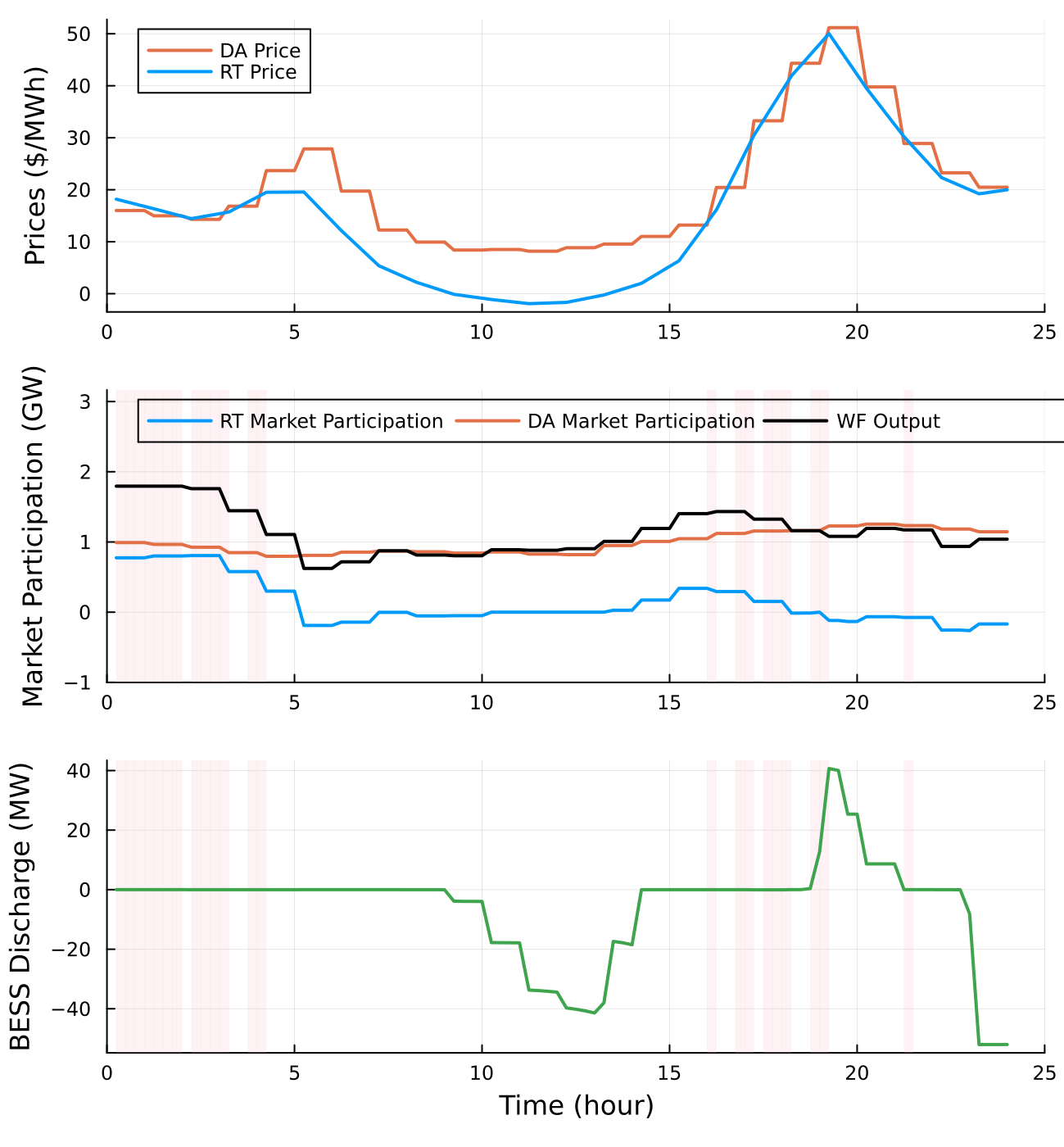}
    \caption{Price, market participation, and ESS operation of COTWDPGE in scenario 80 of case CCD18-3.}
    \label{fig:s80}
\end{figure}

\begin{figure}[ht]
    \centering
    \includegraphics[width=0.55\linewidth]{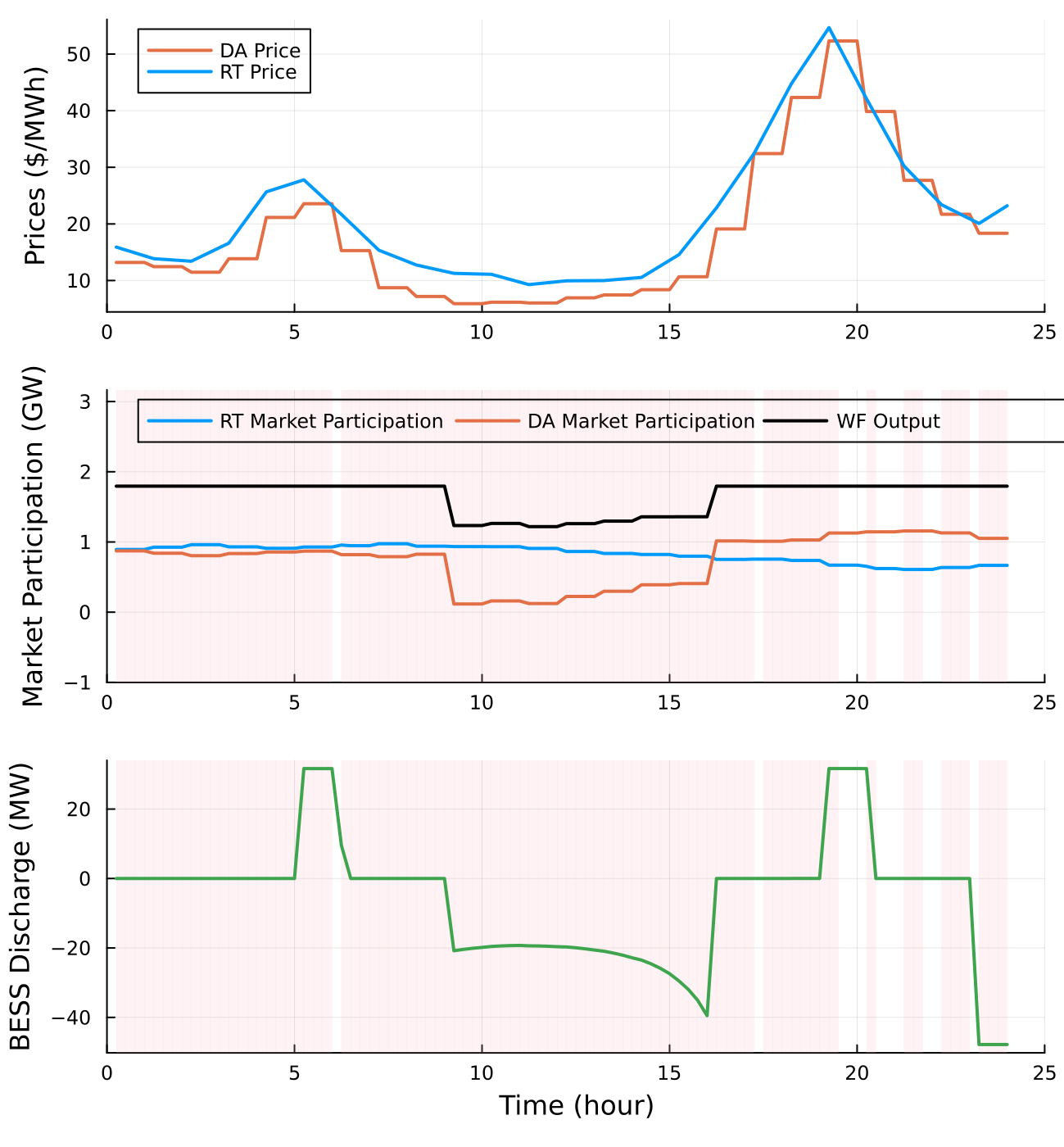}
    \caption{Price, market participation, and ESS operation of COTWDPGE in scenario 83 of case CCD18-3.}
    \label{fig:s83}
\end{figure}

\begin{figure}[ht]
    \centering
    \includegraphics[width=0.55\linewidth]{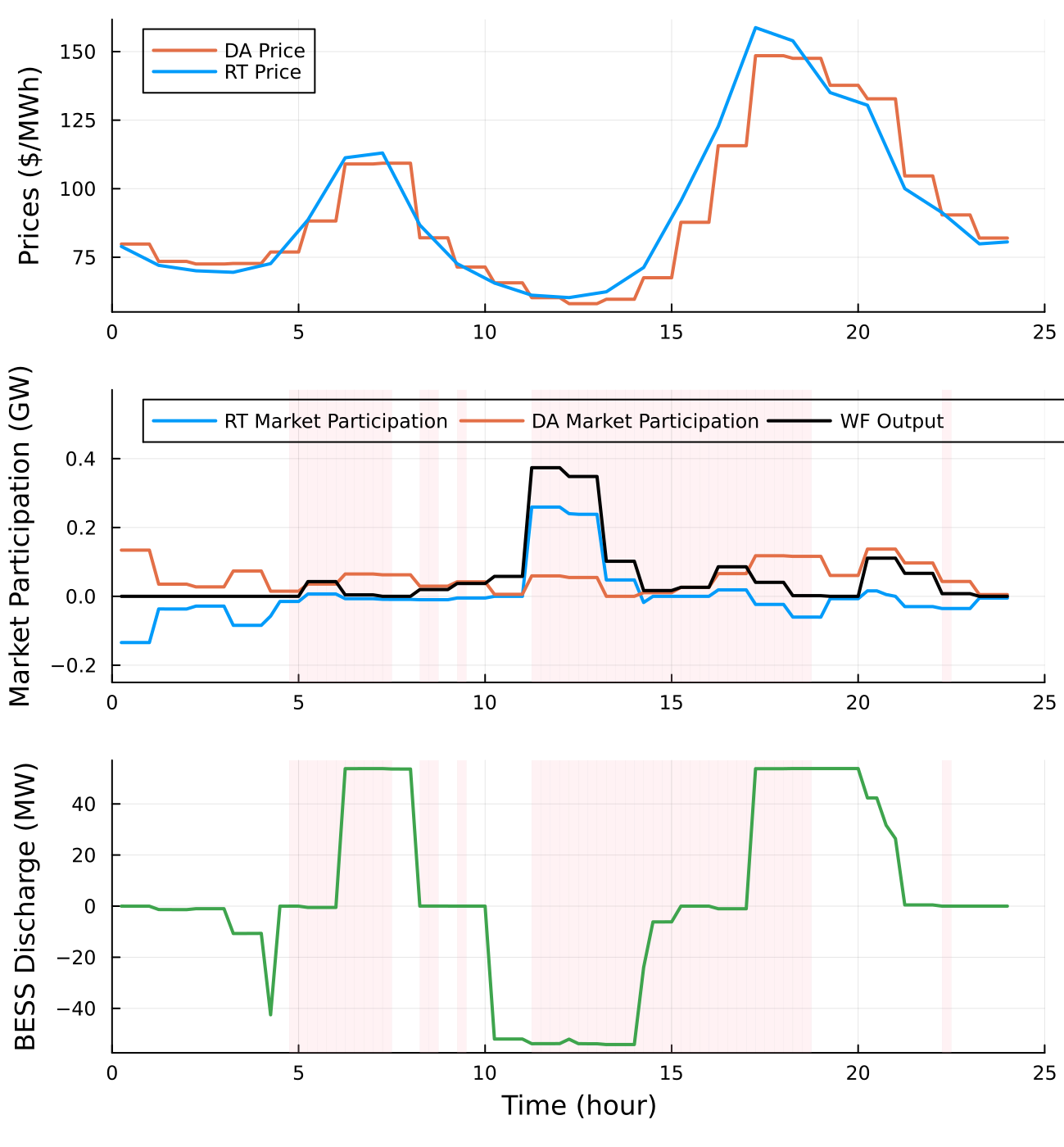}
    \caption{Price, market participation, and ESS operation of COTWDPGE in scenario 99 of case CCD18-3.}
    \label{fig:s99}
\end{figure}

To analyze system operations and market participation strategies, detailed information on three typical scenarios—80, 83, and 99—at the POI COTWDPGE in case CCD18-3 is presented in Figures~\ref{fig:s80}-\ref{fig:s99}. In each figure, the first subfigure displays DA and RT market prices throughout the day. The second subfigure illustrates the wind farm (WF) output and power traded in the two markets, with positive values indicating sales to the markets and negative values indicating purchases from the markets. The final subfigure shows the power discharged from the ESS, where a negative value indicates that the ESS is being charged. In the last two subgraphs, a pink background denotes that the RT market price is higher than the DA market price.

In all scenarios, prices exhibit two peaks: a smaller one around 5:00 am to 7:00 am and a larger one around 5:00 pm to 7:00 pm. The ESS is charged during midday, approximately from 10:00 am to 3:00 pm, and discharged during the two peak periods, particularly the higher peak. The DA market participation decision considers prices and WF output across all scenarios within the same DA stage. Participation in the RT market and joint ESS operation helps reconcile discrepancies between actual WF generation and DA market trades. In scenario 80, the RT price is lower than or close to the DA price, leading to DA market trades nearly matching WF generation and minimal RT market activity. In scenario 83, the RT price exceeds the DA price, with less power allocated to the DA market than WF generation, allowing surplus for the RT market. Even with reduced generation from 9:00 am to 4:00 pm, only DA market trades are decreased, while RT trades remain consistent, and the ESS is charged. In scenario 99, mitigation through the RT market is more pronounced. From 12:00 am to 5:00 am, without WF generation, power is purchased from the RT market to satisfy DA commitments. Between 11:00 am and 1:00 pm, surplus power is sold to the RT market to maintain balance.

\subsection{Revenue Analysis}
Table~\ref{tab_revenue} displays the expected power traded and revenue from DA and RT markets in a day for the four cases with identical data and inflation rates. Average unit revenue is derived by dividing total revenue from DA and RT markets by total power traded. Each case is compared individually with CCD18-3.\

\textbf{Base Case:} This scenario features a smaller ESS size and provides greater reserve support through both the WF and ESS. With the WF holding more reserve, total market-traded power and revenue are reduced. However, similar average unit revenue to CCD18-3 is achieved through ESS operation.\

\textbf{No Reserve Case:} Without needing to reserve WF capacity for frequency support, this scenario trades more power and achieves higher total revenue. As noted in Table~\ref{tab_BESS}, a relatively small ESS is employed for output shifting, resulting in lower average unit revenue.\

\textbf{No ESS Case:} In the absence of ESS, frequency support is solely from the WF, reducing total market-traded power and revenue. RT market trading, without ESS for output shifting, manages the DA commitment and WF generation mismatches, resulting in lower average unit revenue than CCD18-3.

\begin{table}[htbp]
  \centering
  \caption{Power Traded and Revenue in Each Market}
  \resizebox{1\textwidth}{!}{
    \begin{tabular}{|c|c|c|c|c|c|c|c|c|c|c|c|}
    \hline
    \multirow{2}{*}{Case} & \multirow{2}{*}{POI} & \multicolumn{3}{c|}{Traded Power (MW)} & \multicolumn{4}{c|}{Revenue (\$k)} & \multicolumn{3}{c|}{Unit Revenue (\$/MWh)} \\
\cline{3-12}          &       & DA    & RT    & Total & DA    & RT    & Reserve & Total & DA    & RT    & Average \\
    \hline
    \multirow{5}{*}{Base} & WCASCADE & 22362.74 & 482.98 & 22845.72 & 876.80 & 25.58 & 2.69  & 905.07 & 39.21 & 52.96 & 39.50 \\
\cline{2-12}          & John Day & 33008.32 & 983.96 & 33992.29 & 1303.21 & 45.33 & 4.15  & 1352.69 & 39.48 & 46.07 & 39.67 \\
\cline{2-12}          & COTWDPGE & 26506.25 & 627.32 & 27133.57 & 1040.75 & 29.06 & 3.27  & 1073.08 & 39.26 & 46.33 & 39.43 \\
\cline{2-12}          & Tesla & 36533.60 & 1016.89 & 37550.50 & 1422.47 & 49.50 & 4.56  & 1476.53 & 38.94 & 48.68 & 39.20 \\
\cline{2-12}          & Mossland & 23613.67 & 686.54 & 24300.21 & 919.34 & 28.50 & 2.98  & 950.82 & 38.93 & 41.52 & 39.01 \\
    \hline
    \hline
    \multirow{5}{*}{No Reserve} & WCASCADE & 23626.27 & 198.85 & 23825.11 & 922.24 & 14.40 & 0.00  & 936.64 & 39.03 & 72.40 & 39.31 \\
\cline{2-12}          & John Day & 34968.23 & 475.32 & 35443.54 & 1374.49 & 24.11 & 0.00  & 1398.60 & 39.31 & 50.72 & 39.46 \\
\cline{2-12}          & COTWDPGE & 28077.43 & 228.30 & 28305.72 & 1097.70 & 12.84 & 0.00  & 1110.55 & 39.10 & 56.25 & 39.23 \\
\cline{2-12}          & Tesla & 38624.16 & 510.59 & 39134.75 & 1497.64 & 28.39 & 0.00  & 1526.02 & 38.77 & 55.60 & 38.99 \\
\cline{2-12}          & Mossland & 25008.72 & 323.71 & 25332.43 & 969.52 & 13.34 & 0.00  & 982.86 & 38.77 & 41.20 & 38.80 \\
    \hline
    \hline
    \multirow{5}{*}{No ESS} & WCASCADE & 23010.20 & 327.42 & 23337.61 & 899.14 & 19.08 & 1.05  & 919.28 & 39.08 & 58.29 & 39.35 \\
\cline{2-12}          & John Day & 34019.63 & 700.89 & 34720.51 & 1338.57 & 32.86 & 1.62  & 1373.05 & 39.35 & 46.88 & 39.50 \\
\cline{2-12}          & COTWDPGE & 27318.77 & 403.52 & 27722.29 & 1068.92 & 19.66 & 1.28  & 1089.87 & 39.13 & 48.73 & 39.27 \\
\cline{2-12}          & Tesla & 37613.02 & 730.73 & 38343.76 & 1459.97 & 36.67 & 1.76  & 1498.40 & 38.82 & 50.18 & 39.03 \\
\cline{2-12}          & Mossland & 24339.10 & 478.40 & 24817.50 & 944.29 & 19.42 & 1.14  & 964.86 & 38.80 & 40.59 & 38.83 \\
    \hline
    \hline
    \multirow{5}{*}{CCD18-3} & WCASCADE & 23326.19 & 293.12 & 23619.31 & 914.05 & 19.10 & 1.30  & 934.45 & 39.19 & 65.16 & 39.51 \\
\cline{2-12}          & John Day & 34480.43 & 657.19 & 35137.62 & 1360.74 & 33.64 & 2.05  & 1396.42 & 39.46 & 51.19 & 39.68 \\
\cline{2-12}          & COTWDPGE & 27692.32 & 363.91 & 28056.22 & 1087.48 & 19.09 & 1.59  & 1108.16 & 39.27 & 52.45 & 39.44 \\
\cline{2-12}          & Tesla & 38061.13 & 734.58 & 38795.71 & 1482.51 & 39.05 & 2.28  & 1523.85 & 38.95 & 53.16 & 39.22 \\
\cline{2-12}          & Mossland & 24648.57 & 465.65 & 25114.22 & 960.07 & 20.17 & 1.52  & 981.76 & 38.95 & 43.32 & 39.03 \\
    \hline
    \end{tabular}}%
  \label{tab_revenue}%
\end{table}%

\subsection{Validation of control co-design offshore wind farm using simulations (PSCAD)}

\begin{figure}[ht]
    \centering
    \includegraphics[width=0.49\linewidth]{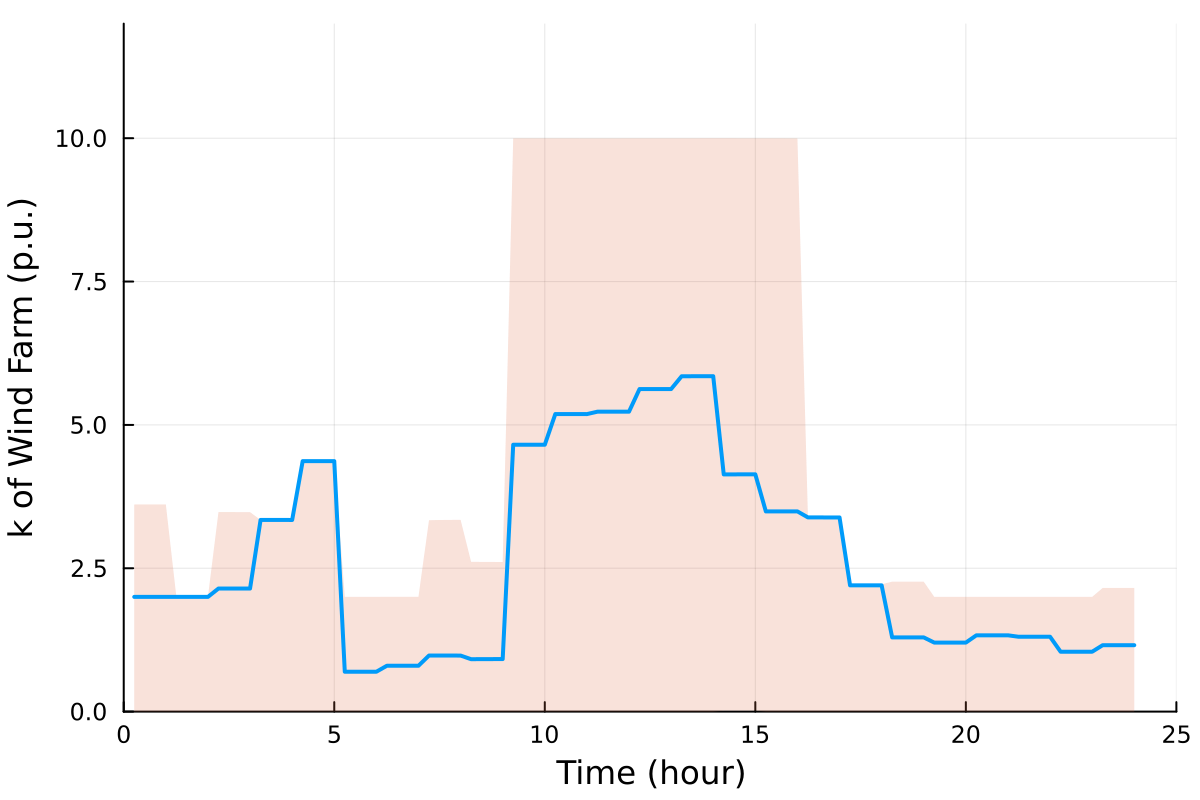}
    \includegraphics[width=0.49\linewidth]{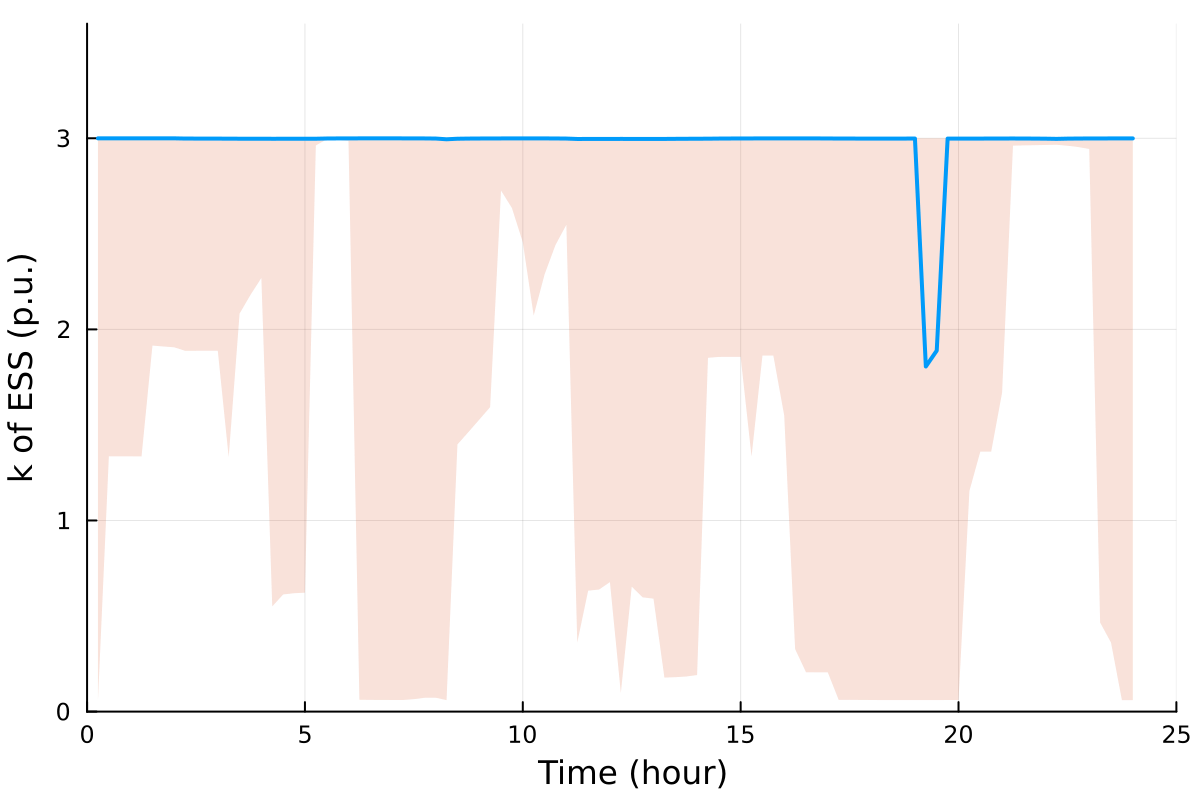}
    \caption{Droop control variable $k$ at COTWDPGE on scenario 80 in case CCD18-3.}
    \label{fig:ks99}
\end{figure}

In Figure~\ref{fig:ks99}, we present the plot of the droop control variable $k$ at COTWDPGE for scenario 80 in case CCD18-3. The pink band illustrates the distribution of this variable across all scenarios. For most of the time, the ESS variable is at its upper limit. By adjusting the droop status of the wind farm and ESS, the system optimizes its output while ensuring sufficient reserve for frequency support.

\begin{figure}[htbp]
   \centering
   \subfloat[]{
        \includegraphics[scale=0.35]{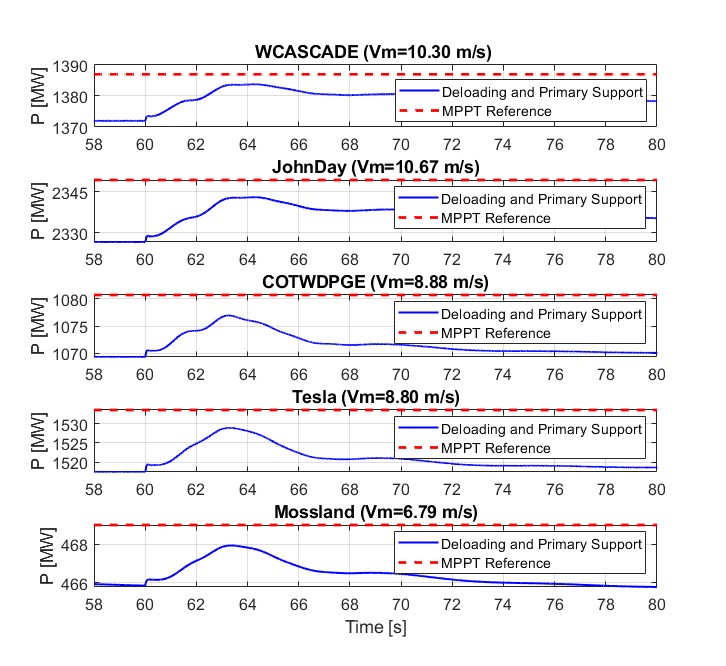}
        }
%        \vspace{-0em} 
   \hfill
   \subfloat[]{
         \includegraphics[scale=0.36]{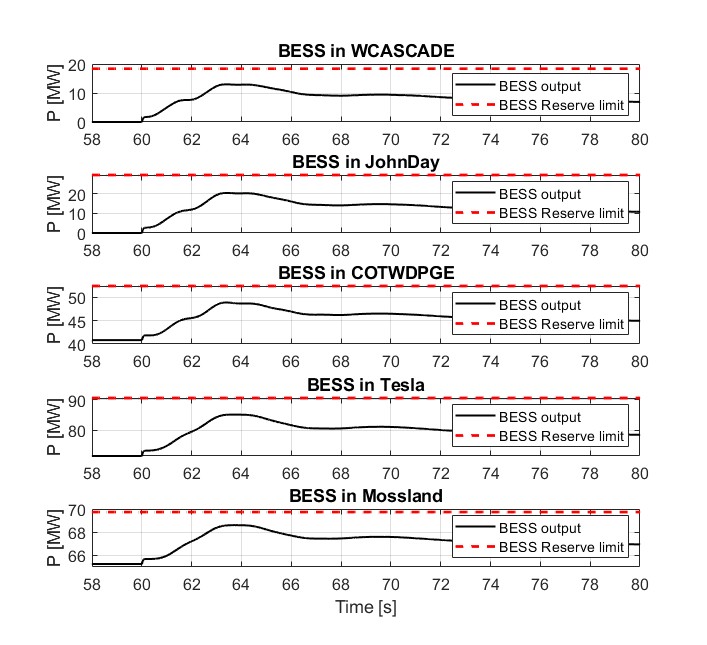}
         }   
\caption{Power curves of 240-bus miniWECC system in PSCAD: (a) OWF; (b): BESS}
\label{fig:pscad_power}
\end{figure}

\begin{figure}[htbp]
   \centering
   \subfloat[]{
         \includegraphics[scale=0.5]{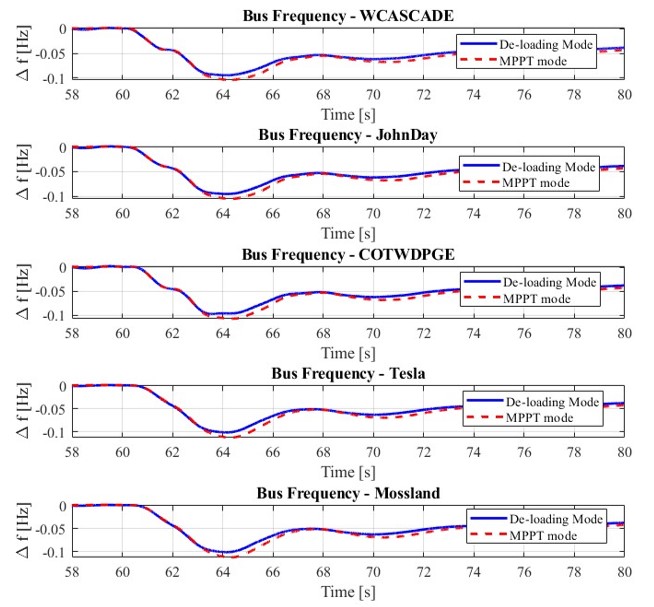}
         }
\caption{POI frequency curves of 240-bus miniWECC system in PSCAD.}
\label{fig:pscad_frequency}
\end{figure}

To further validate the dynamic performance of the co-design results, a time-domain electromagnetic transient (EMT) simulation was conducted using PSCAD \cite{she2024configuration}. Both Maximum Power Point Tracking (MPPT) and de-loading control strategies were integrated and tested under the scenario of losing the largest generator at Palo Verde, Arizona, at 60 seconds, which had an output of 2182.3 MW before tripping. Additional details on the PSCAD model can be found in \ref{apd:pscad}. Transient trajectories of power and frequency, depicted in Figures \ref{fig:pscad_power} and \ref{fig:pscad_frequency}, were then analyzed to verify the stability and effectiveness of the co-design results.

Overall, the miniWECC system integrating OWFs remains stable both before and after a generator trip on the main grid. Prior to the 60-second mark, the steady-state output of the four OWFs shows only minor differences compared to the co-design results. After the disturbance, the power output and the increase in power are both smaller than those of the MPPT curves, indicating sufficient headroom. Additionally, the frequency nadir is lower when the OWF operates in MPPT mode without providing frequency support. Consequently, the co-design approach demonstrates superior transient performance, which is crucial given the increasing penetration of wind and other inverter-based resources.

% \subsection{Role of energy storage with Offshore wind farms}
% \HS{HS, WW: To complete the section}
% \subsection{Impact of inflation's rate on Offshore wind farm revenue}
% \HS{HS, WW: To complete the section}
% \subsection{Impact of pricing data on design and operation of offshore wind farm}
% \HS{HS, WW: To complete the section}

% \pagebreak

% \section{Discussion}
% \HS{HS, WW: To complete the section}\\

% Discuss the limitation of the current framework 
\section{Conclusions and Future Work}\label{sec:conc}
Simultaneous consideration of operations, controls, and market participation during the planning of renewable energy systems, such as offshore wind farms, can achieve long-term reliability and optimal performance. Incorporating various uncertainties into the decision-making process enhances overall system resilience. In this work, we develop and present a multi-timescale, multi-stage stochastic control co-design methodology. This approach models the multi-timescale nature of energy market participation and addresses uncertainties associated with renewable resources and prices at the interconnection point for revenue maximization. The approach is demonstrated using an offshore wind farm grid integration case study. Here, we optimize the sizing of HVDC power export cables from offshore to onshore to prevent under-sizing and over-sizing, which could significantly impact project investment costs. Furthermore, we consider energy storage sizing, addressing grid stabilization needs and incorporating market participation. Both offshore wind farm and energy storage participation in ancillary service markets are evaluated for additional revenue opportunities for wind farm owners. Our control co-design approach optimizes cable and energy storage sizing, and market participation with droop parameters, showing improved annual revenue for five different capacity wind farms studied. A scenario tree-based stochastic formulation incorporates uncertainty. Additionally, the framework allows evaluation of various inflation rates and energy storage implications on overall revenue. The study provides validation for droop control parameters through transient PSSE simulations of offshore wind farm integrated with the miniWECC AC grid model, showing comparable reserve values and performance in meeting the integrated power grid's frequency requirements.

Currently, we rely on historical wind speed and price data, as offshore wind farm development in the U.S. is still maturing. Future efforts could enhance price data estimation using detailed production cost planning with offshore wind farms and consider futuristic climate scenarios for wind speed data. Furthermore, different network topologies (e.g., meshed) could be explored for power wheeling, with incentive designs providing additional revenue streams for offshore wind farm owners. The study can also expand by including additional substation-level design and control variables and incorporating reduced-order models of dynamic converter-inverters. As energy market participation becomes increasingly important for future renewable energy developers, the proposed framework could be extended to next-generation energy market models to estimate revenue and develop participation strategies.

% Limitation;
% 1. As we don't have the actual production data for the wind farm we use the wind-turbine power curve and upscale it. While this is a crude estimate of reality in future the study will consider detailed model of offshore wind power production which incorporate wake-effects and additional phenomena that could affect the annual power production from the offshore wind farm.
% 2. Addtionally, we could consider the topologies of interconnections that can allow for generating revenue as power wheeling services to additionally  the t
% 3. Considering optimization/sizing of substations and exploring possiblibilties of providing grid frequency support through capacitance based (VSC-HVDC) c

\section*{Acknowledgment}
This research was supported by the Energy System Co-Design with Multiple Objectives and Power Electronics (E-COMP) Initiative, under the Laboratory Directed Research and Development (LDRD) Program at Pacific Northwest National Laboratory (PNNL). PNNL is a multi-program national laboratory operated for the U.S. Department of Energy (DOE) by Battelle Memorial Institute under Contract No. DE-AC05-76RL01830. The authors express their gratitude to Jay Barlow for his invaluable discussions regarding the offshore wind farm use case and his assistance in identifying relevant datasets for the case study.

% Declaration of generative AI and AI-assisted technologies in the writing process

% During the preparation of this work the author(s) used [NAME TOOL / SERVICE] in order to [REASON]. After using this tool/service, the author(s) reviewed and edited the content as needed and take(s) full responsibility for the content of the publication.

\appendix
\section{Offshore Wind Farm Turbine Technology}
% Power curve data
% anyother specifics.

% \HS{HS, Wei: Describe the turbine technology used for the study}

\begin{figure}[t]
    \centering
    \includegraphics[width=0.7\linewidth]{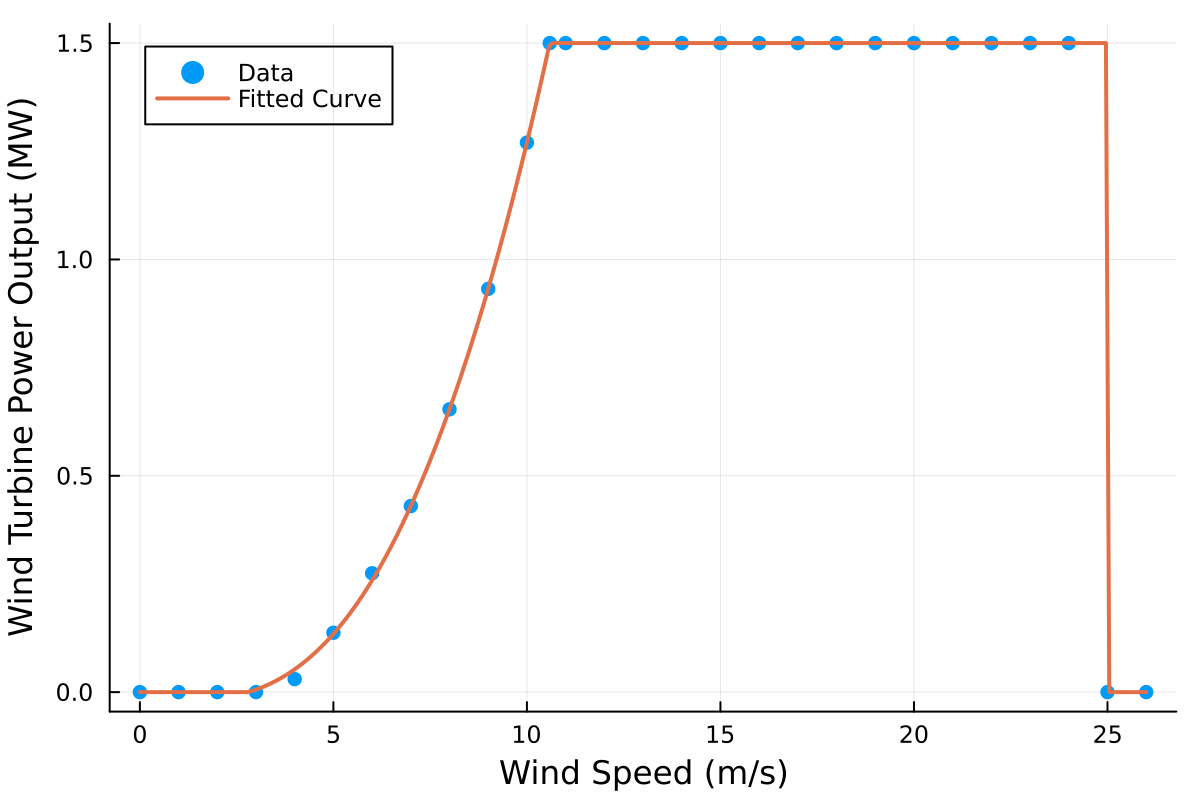}
    \caption{Wind Turbine Power Curve}
    \label{fig:WTPC}
\end{figure}

We use the GE 1.5 MW wind turbine as a base model \cite{miller2003dynamic} to generate sampling points for speed-power fitting. A detailed electromagnetic transient (EMT) model was developed in PSCAD, and simulations were conducted by incrementally increasing the wind speed from the cut-in to the cut-out wind speed. These points are then fitted with a piece-wise quadratic function as presented in Figure~\ref{fig:WTPC}. The overall fitted curve is aligned with the original GE 1.5 MW wind turbine \cite{miller2003dynamic}. We upscale the fitted power curve to the rated power of each OWF to convert wind speed to power output in the control co-design process.

\section{Wind Data} \label{app1}
The wind data utilized in the study includes historical wind speed data available from various sources, particularly focusing on offshore wind areas. The dataset includes wind speed measurements at different heights (e.g., 140m) and covers a range of years, notably from 2000 to 2020. The data is essential for generating stochastic wind scenarios and validating wind speed distributions using methods such as the Weibull distribution. Key sources for this wind data include the NREL WIND Toolkit Offshore Summary Dataset, which offers data for specific years such as 2017 and 2020.

The Offshore California Dataset from \href{https://developer.nrel.gov/docs/wind/wind-toolkit/}{NREL WIND Toolkit} is a 21-year wind resource dataset for offshore California. Produced in 2020, this data set replaces NREL's Wind Integration National Dataset (WIND) Toolkit for offshore California, which was produced and released publicly in 2013 and is currently the principal data set used by stakeholders for wind resource assessment in the continental United States. Both the WIND Toolkit and this new data set are created using the Weather Research and Forecasting (WRF) numerical weather prediction model (NWP)

\begin{figure}[htp]
    \centering
    \includegraphics[width=0.8\linewidth]{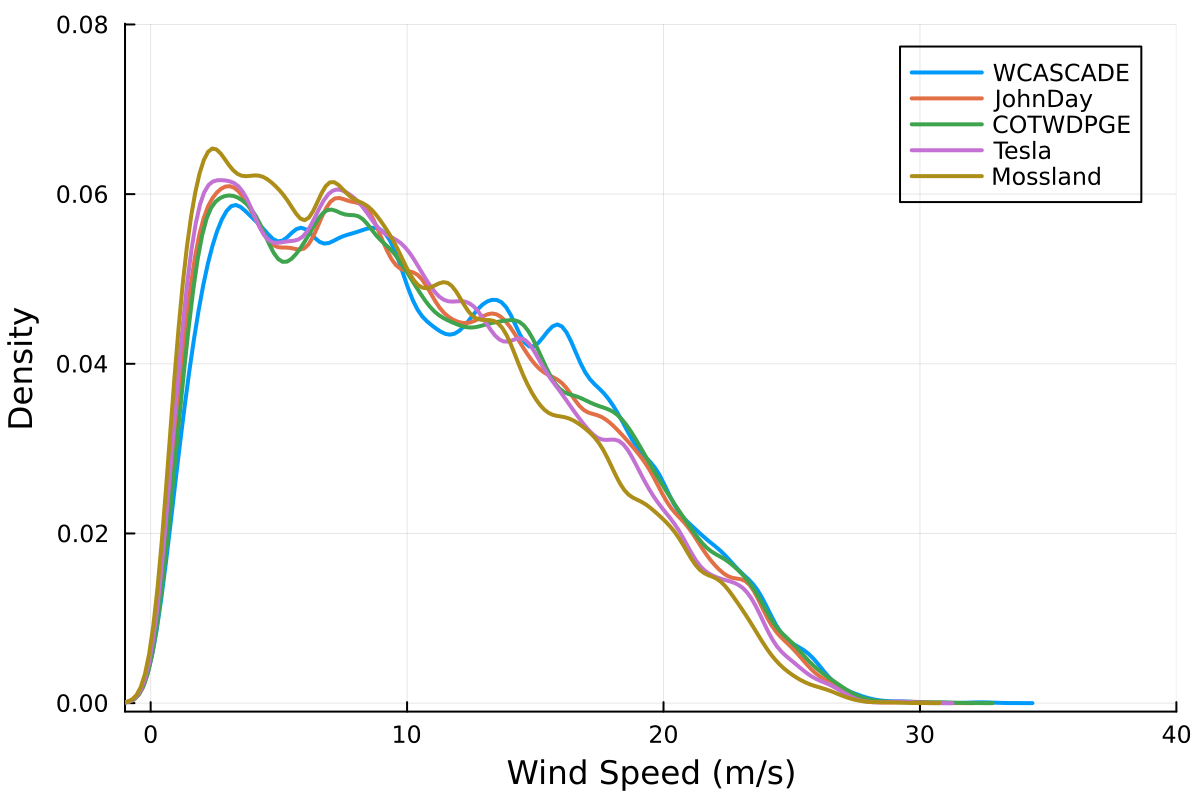}
    \caption{Density of Wind Speed in 2018}
    \label{fig:wind18}
\end{figure}

\begin{figure}[htp]
    \centering
    \includegraphics[width=0.8\linewidth]{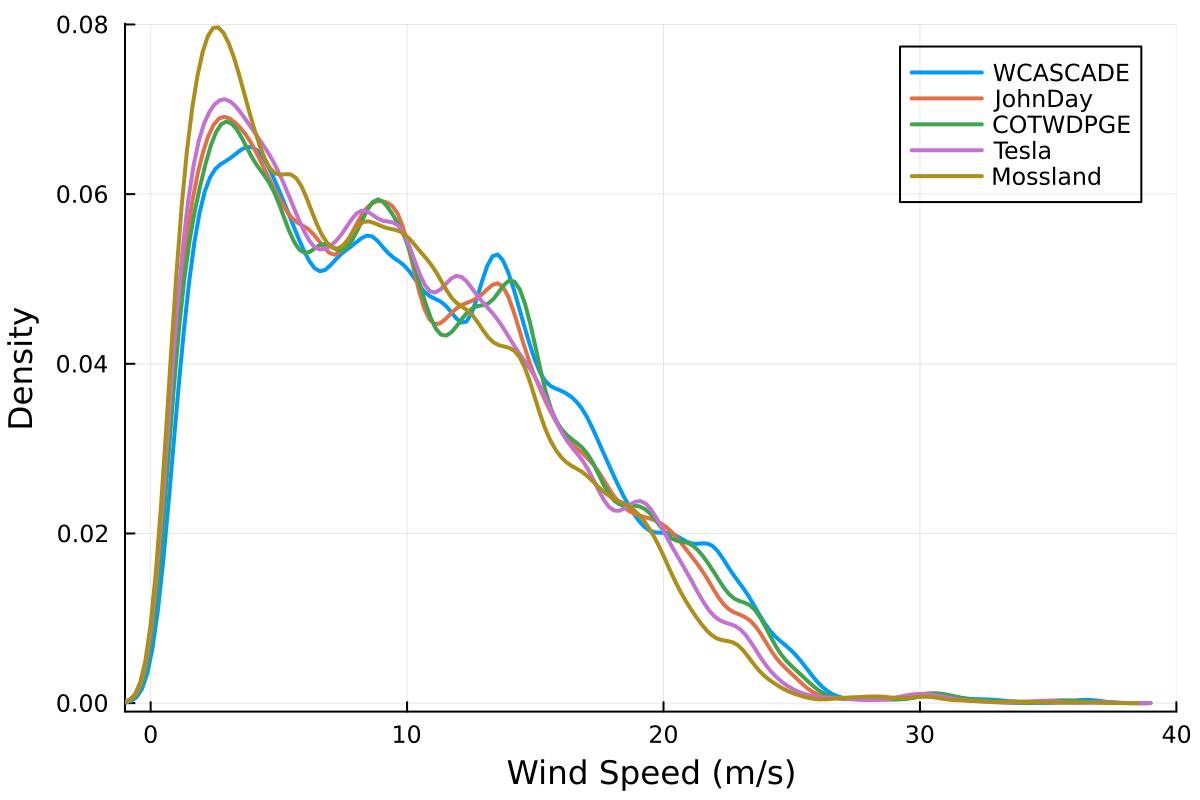}
    \caption{Density of Wind Speed in 2022}
    \label{fig:wind22}
\end{figure}

The Figure ~\ref{fig:wind18}-\ref{fig:wind22} illustrate wind speed density distributions at various locations (WCASCADE, JohnDay, COTWDPGE, Tesla, and Mossland) for 2018 and 2022. Both years show multi-modal distributions, with peaks around 4-10 m/s and tailing off after 30 m/s. In 2022, the density appears slightly higher around 8-12 m/s across most locations compared to 2018, which might indicate stronger winds or more frequent occurrences of mid-range speeds. This shift can significantly influence wind power generation since wind turbines typically produce maximum power output in this speed range.

Comparing the wind speed distributions to the turbine power curve in Figure~\ref{fig:WTPC}, the observed wind speeds in 2018 and 2022 indicate that a significant portion of the wind speeds in both years fall within the optimal power generation region (8-12 m/s). The slight increase in frequency in this range during 2022 suggests potentially higher energy output, assuming the wind farms are designed for maximum efficiency in this range.

\section{Energy Market Data} \label{app2}

The energy market data primarily comes from the California Independent System Operator (CAISO). 
This data is accessed through two sources, the CAISO OASIS API or EIA portal. 

CAISO OASIS API includes various market-related metrics such as Locational Marginal Prices (LMPs) for both day-ahead and real-time markets, Ancillary Service Prices (ASP), and system performance data like Regulation Mileage. The CAISO data covers the periods from 2018 onwards, with some limitations in earlier years where LMP data may not be available. This data is crucial for analyzing the market participation of offshore wind farms and optimizing their dispatch in energy markets.

The Wholesale Electricity Market Portal, launched by the U.S. Energy Information Administration (EIA) in March 2024, provides access to electricity market data from seven Regional Transmission Organizations (RTOs) and Independent System Operators (ISOs), including CAISO. The Portal offers datasets such as day-ahead and real-time locational marginal prices (LMPs), load/demand data, generation fuel mix, and city temperatures, gathered from RTO/ISO public data and NOAA. Data availability varies, with updates ranging from hourly to longer intervals. Data can be accessed through dashboards and bulk download flat files via \href{https://www.eia.gov/electricity/wholesalemarkets/data.php?rto=caiso}{CAISO EIA Portal}, though API access is not currently available.

\section{Scenario Generation Algorithm}\label{apd:scenario}
 The scenario generation algorithm is presented in Algorithm-\ref{alg:kde}.
\begin{algorithm}[h]
\caption{Kernel-Based Scenario Tree Generation Algorithm}\label{alg:kde}
\begin{algorithmic}[1]
\State \textbf{Input:} Time series data for LMP, ASP, Windspeed, etc.
\State \textbf{Parameters:} $SCN1 = 20$ (second-stage scenarios), $SCN2 = 5$ (third-stage scenarios)

\State \textbf{Step 1: Data Preprocessing}
\State Load datasets for LMP, ASP, and Windspeed
\State Reshape data into a 3D array $(N, T, K)$, where $N$ is the number of samples, $T$ is time steps, and $K$ is the number of variables

\State \textbf{Step 2: Scenario Generation Using Kernel Density Estimation (KDE)}
\For{each time step $t = 1$ to $T$}
    \State Normalize weights $w \gets w / \sum w$
    \State Compute effective sample size $N_t$ and standard deviation $\sigma_t$
    \State Generate scenario $x[t]$ using KDE from multivariate normal distribution
    \State Update weights based on Markovian or Non-Markovian process
\EndFor
\State Return scenario trajectory $x$

\State \textbf{Step 3: Tree Construction}
\State Define tree structure as $[1, SCN1, SCN2, 1, 1, 1]$
\State Apply kernel-based sampling to generate $100,000$ scenarios
\State Construct the scenario tree using tree approximation

\State \textbf{Step 4: Save the Scenario Tree}

\end{algorithmic}
\end{algorithm}

\section{PSCAD Model and Test Scenarios}\label{apd:pscad}
The PSCAD model used for validation is modified from the base 240-bus miniWECC system developed in \cite{she2024configuration}. Grid-following inverter-based resources with dynamic models of REGCA1, REECB1, and REPCA1 are integrated into the system. The REGCA1 model is adjusted so that the \textit{d}- and \textit{q}-axis currents are transformed into a three-phase current using a phase-locked loop. Additionally, frequency and voltage droop deadbands are set at ±0.017 Hz and ±0.01 p.u., respectively.

Five offshore wind farms (OWFs) are connected to WCASCADE, JOHN DAY, COTTONWOOD, TESLA, and MOSSLAND, with rated capacities of 1500 MW, 2350 MW, 1810 MW, 2640 MW, and 1800 MW, respectively. These type 3 turbines include a pitch angle control module, a grid-side control module, a rotor-side control module, and a DC-link chopper, allowing them to switch flexibly between MPPT and de-loading modes. They are assigned wind speeds of 10.68 m/s, 10.68 m/s, 8.86 m/s, 8.86 m/s, and 6.75 m/s, representing typical wind conditions.

\bibliographystyle{elsarticle-num} 
\bibliography{ifacconf}

\end{document}